\pdfoutput=1
\documentclass[12pt,a4paper]{article}

\usepackage{ifthen} 
\newboolean{pdflatex}
\setboolean{pdflatex}{true} 

\newboolean{articletitles}
\setboolean{articletitles}{true} 

\newboolean{uprightparticles}
\setboolean{uprightparticles}{false} 

\def\paperauthors{LHCb collaboration} 

\def\paperasciititle{Measurement of the ratios of branching fractions R(D*) and R(D^0)} 

\def\papertitle{Measurement of the ratios of branching fractions \RDst and \RDz} 

\def\paperkeywords{{High Energy Physics}, {LHCb}} 

\def\papercopyright{\the\year\ CERN for the benefit of the LHCb collaboration} 
\def\paperlicence{CC BY 4.0 licence}
\def\paperlicenceurl{https://creativecommons.org/licenses/by/4.0/}

\usepackage[top=1in, bottom=1.25in, left=1in, right=1in]{geometry}

\usepackage{microtype}
\usepackage{lineno}  
\usepackage{xspace} 
\usepackage{caption} 

\usepackage{graphicx}  
\usepackage{color}
\usepackage{colortbl}
\graphicspath{{./figs/}} 

\usepackage{amsmath} 
\usepackage{amssymb}
\usepackage{amsfonts}
\usepackage{upgreek} 

\newcommand*\patchAmsMathEnvironmentForLineno[1]{%
\expandafter\let\csname old#1\expandafter\endcsname\csname #1\endcsname
\expandafter\let\csname oldend#1\expandafter\endcsname\csname
end#1\endcsname
 \renewenvironment{#1}%
   {\linenomath\csname old#1\endcsname}%
   {\csname oldend#1\endcsname\endlinenomath}%
}
\newcommand*\patchBothAmsMathEnvironmentsForLineno[1]{%
  \patchAmsMathEnvironmentForLineno{#1}%
  \patchAmsMathEnvironmentForLineno{#1*}%
}
\AtBeginDocument{%
\patchBothAmsMathEnvironmentsForLineno{equation}%
\patchBothAmsMathEnvironmentsForLineno{align}%
\patchBothAmsMathEnvironmentsForLineno{flalign}%
\patchBothAmsMathEnvironmentsForLineno{alignat}%
\patchBothAmsMathEnvironmentsForLineno{gather}%
\patchBothAmsMathEnvironmentsForLineno{multline}%
\patchBothAmsMathEnvironmentsForLineno{eqnarray}%
}

\usepackage{hyperxmp}

\usepackage[pdftex,
            pdfauthor={\paperauthors},
            pdftitle={\paperasciititle},
            pdfkeywords={\paperkeywords},
            pdfcopyright={Copyright (C) \papercopyright},
            pdflicenseurl={\paperlicenceurl}]{hyperref}

\usepackage[bottom,flushmargin,hang,multiple]{footmisc}

\usepackage[all]{hypcap} 

\usepackage{xspace} 
\usepackage{upgreek}

\def\lhcb   {\mbox{LHCb}\xspace}

\def\MagUp {\mbox{\em Mag\kern -0.05em Up}\xspace}

\ifthenelse{\boolean{uprightparticles}}%
{

 \def\Pmu         {\ensuremath{\upmu}\xspace}                 
 \def\Pnu         {\ensuremath{\upnu}\xspace}                 
                  
 \def\Ppi         {\ensuremath{\uppi}\xspace}

 \def\Ptau        {\ensuremath{\uptau}\xspace}

 \def\Ppsi        {\ensuremath{\uppsi}\xspace}

 \def\PDelta      {\ensuremath{\Delta}\xspace}                 
 \def\PXi         {\ensuremath{\Xi}\xspace}                 
 \def\PLambda     {\ensuremath{\Lambda}\xspace}                 
 \def\PSigma      {\ensuremath{\Sigma}\xspace}                 
 \def\POmega      {\ensuremath{\Omega}\xspace}                 
 \def\PUpsilon    {\ensuremath{\Upsilon}\xspace}
 \let\oldPi\Pi
 \def\PPi         {\ensuremath{\oldPi}\xspace}

 \def\PB      {\ensuremath{\mathrm{B}}\xspace}                 
                  
 \def\PD      {\ensuremath{\mathrm{D}}\xspace}

 \def\PJ      {\ensuremath{\mathrm{J}}\xspace}                 
 \def\PK      {\ensuremath{\mathrm{K}}\xspace}

 \def\Pb      {\ensuremath{\mathrm{b}}\xspace}                 
 \def\Pc      {\ensuremath{\mathrm{c}}\xspace}                 
                  
 \def\Pe      {\ensuremath{\mathrm{e}}\xspace}

 \def\Pi      {\ensuremath{\mathrm{i}}\xspace}

 \def\Pp      {\ensuremath{\mathrm{p}}\xspace}

 \def\Ps      {\ensuremath{\mathrm{s}}\xspace}

 \def\thebaroffset{0.0em}
}
{

 \def\Pmu         {\ensuremath{\mu}\xspace}                 
 \def\Pnu         {\ensuremath{\nu}\xspace}                 
                  
 \def\Ppi         {\ensuremath{\pi}\xspace}

 \def\Ptau        {\ensuremath{\tau}\xspace}

 \def\Ppsi        {\ensuremath{\psi}\xspace}                 
                  
 \mathchardef\PDelta="7101
 \mathchardef\PXi="7104
 \mathchardef\PLambda="7103
 \mathchardef\PSigma="7106
 \mathchardef\POmega="710A
 \mathchardef\PUpsilon="7107
 \mathchardef\PPi="7105
                  
 \def\PB      {\ensuremath{B}\xspace}                 
                  
 \def\PD      {\ensuremath{D}\xspace}

 \def\PJ      {\ensuremath{J}\xspace}                 
 \def\PK      {\ensuremath{K}\xspace}

 \def\Pb      {\ensuremath{b}\xspace}                 
 \def\Pc      {\ensuremath{c}\xspace}                 
                  
 \def\Pe      {\ensuremath{e}\xspace}

 \def\Pi      {\ensuremath{i}\xspace}

 \def\Pp      {\ensuremath{p}\xspace}

 \def\Ps      {\ensuremath{s}\xspace}

 \def\thebaroffset{0.18em}
}
\newcommand{\offsetoverline}[2][\thebaroffset]{\kern #1\overline{\kern -#1 #2}}%

\makeatletter
\ifcase \@ptsize \relax
  \newcommand{\miniscule}{\@setfontsize\miniscule{4}{5}}
\or
  \newcommand{\miniscule}{\@setfontsize\miniscule{5}{6}}
\or
  \newcommand{\miniscule}{\@setfontsize\miniscule{5}{6}}
\fi
\makeatother

\DeclareRobustCommand{\optbar}[1]{\shortstack{{\miniscule (\rule[.5ex]{1.25em}{.18mm})}
  \\ [-.7ex] $#1$}}

\def\electron   {{\ensuremath{\Pe}}\xspace}

\def\mup        {{\ensuremath{\Pmu^+}}\xspace}
\def\mun        {{\ensuremath{\Pmu^-}}\xspace} 

\def\taum       {{\ensuremath{\Ptau^-}}\xspace}

\def\ellm       {{\ensuremath{\ell^-}}\xspace}

\def\neu        {{\ensuremath{\Pnu}}\xspace}
\def\neub       {{\ensuremath{\overline{\Pnu}}}\xspace}
\def\neum       {{\ensuremath{\neu_\mu}}\xspace}
\def\neumb      {{\ensuremath{\neub_\mu}}\xspace}
\def\neut       {{\ensuremath{\neu_\tau}}\xspace}
\def\neutb      {{\ensuremath{\neub_\tau}}\xspace}

\def\neulb      {{\ensuremath{\neub_\ell}}\xspace}

\def\squark    {{\ensuremath{\Ps}}\xspace}

\def\cquark    {{\ensuremath{\Pc}}\xspace}

\def\bquark    {{\ensuremath{\Pb}}\xspace}

\def\pion   {{\ensuremath{\Ppi}}\xspace}
\def\piz    {{\ensuremath{\pion^0}}\xspace}
\def\pip    {{\ensuremath{\pion^+}}\xspace}
\def\pim    {{\ensuremath{\pion^-}}\xspace}

\def\kaon    {{\ensuremath{\PK}}\xspace}

\def\KorKbar {\kern \thebaroffset\optbar{\kern -\thebaroffset \PK}{}\xspace}

\def\Kp      {{\ensuremath{\kaon^+}}\xspace}
\def\Km      {{\ensuremath{\kaon^-}}\xspace}

\def\D       {{\ensuremath{\PD}}\xspace}

\def\DorDbar {\kern \thebaroffset\optbar{\kern -\thebaroffset \PD}\xspace}
\def\Dz      {{\ensuremath{\D^0}}\xspace}

\def\Dp      {{\ensuremath{\D^+}}\xspace}
\def\Dm      {{\ensuremath{\D^-}}\xspace}

\def\DpDm    {\ensuremath{\Dp {\kern -0.16em \Dm}}\xspace}
\def\Dstar   {{\ensuremath{\D^*}}\xspace}

\def\Dstarz  {{\ensuremath{\D^{*0}}}\xspace}

\def\Dstarp  {{\ensuremath{\D^{*+}}}\xspace}

\def\B       {{\ensuremath{\PB}}\xspace}
\def\Bbar    {{\ensuremath{\offsetoverline{\PB}}}\xspace}
\def\Bb      {{\ensuremath{\Bbar}}\xspace}
\def\BorBbar {\kern \thebaroffset\optbar{\kern -\thebaroffset \PB}\xspace}

\def\Bzb     {{\ensuremath{\Bbar{}^0}}\xspace}
\def\Bd      {{\ensuremath{\B^0}}\xspace}

\def\BdorBdbar {\kern \thebaroffset\optbar{\kern -\thebaroffset \Bd}\xspace}
\def\Bu      {{\ensuremath{\B^+}}\xspace}
\def\Bub     {{\ensuremath{\B^-}}\xspace}
\def\Bp      {{\ensuremath{\Bu}}\xspace}
\def\Bm      {{\ensuremath{\Bub}}\xspace}

\def\Bs      {{\ensuremath{\B^0_\squark}}\xspace}
\def\Bsb     {{\ensuremath{\Bbar{}^0_\squark}}\xspace}
\def\BsorBsbar {\kern \thebaroffset\optbar{\kern -\thebaroffset \Bs}\xspace}

\def\Bcm     {{\ensuremath{\B_\cquark^-}}\xspace}

\def\jpsi     {{\ensuremath{{\PJ\mskip -3mu/\mskip -2mu\Ppsi}}}\xspace}

\def\Y#1S{\ensuremath{\PUpsilon{(#1S)}}\xspace}

\def\proton      {{\ensuremath{\Pp}}\xspace}

\def\Lz          {{\ensuremath{\PLambda}}\xspace}

\def\LorLbar     {\kern \thebaroffset\optbar{\kern -\thebaroffset \PLambda}\xspace}

\def\Lc          {{\ensuremath{\Lz^+_\cquark}}\xspace}

\def\Lb           {{\ensuremath{\Lz^0_\bquark}}\xspace}

\def\BF         {{\ensuremath{\mathcal{B}}}\xspace}
\def\BR         {\BF}

\newcommand{\decay}[2]{\ensuremath{#1\!\to #2}\xspace} 

\def\to                 {\ensuremath{\rightarrow}\xspace}

\def\AT#1     {\ensuremath{A_{\mathrm{T}}^{#1}}\xspace}           

\def\C#1      {\ensuremath{\mathcal{C}_{#1}}\xspace}                       
\def\Cp#1     {\ensuremath{\mathcal{C}_{#1}^{'}}\xspace}                    
\def\Ceff#1   {\ensuremath{\mathcal{C}_{#1}^{\mathrm{(eff)}}}\xspace}        
\def\Cpeff#1  {\ensuremath{\mathcal{C}_{#1}^{'\mathrm{(eff)}}}\xspace}       
\def\Ope#1    {\ensuremath{\mathcal{O}_{#1}}\xspace}                       
\def\Opep#1   {\ensuremath{\mathcal{O}_{#1}^{'}}\xspace}                    

\newcommand{\aunit}[1]{\ensuremath{\text{\,#1}}}       

\newcommand{\tev}{\aunit{Te\kern -0.1em V}\xspace}
\newcommand{\gev}{\aunit{Ge\kern -0.1em V}\xspace}
\newcommand{\mev}{\aunit{Me\kern -0.1em V}\xspace}
\newcommand{\kev}{\aunit{ke\kern -0.1em V}\xspace}
\newcommand{\ev}{\aunit{e\kern -0.1em V}\xspace}
 
\newcommand{\mevc}{\ensuremath{\aunit{Me\kern -0.1em V\!/}c}\xspace}
\newcommand{\gevc}{\ensuremath{\aunit{Ge\kern -0.1em V\!/}c}\xspace}
\newcommand{\mevcc}{\ensuremath{\aunit{Me\kern -0.1em V\!/}c^2}\xspace}
\newcommand{\gevcc}{\ensuremath{\aunit{Ge\kern -0.1em V\!/}c^2}\xspace}
\newcommand{\gevgevcccc}{\ensuremath{\gev^2\!/c^4}\xspace} 

\def\fb   {\ensuremath{\aunit{fb}}\xspace}
\def\invfb   {\ensuremath{\fb^{-1}}\xspace}

\def\gsim{{~\raise.15em\hbox{$>$}\kern-.85em
          \lower.35em\hbox{$\sim$}~}\xspace}
\def\lsim{{~\raise.15em\hbox{$<$}\kern-.85em
          \lower.35em\hbox{$\sim$}~}\xspace}

\def\pt         {\ensuremath{p_{\mathrm{T}}}\xspace}

\def\photos     {\mbox{\textsc{Photos}}\xspace}

\def\roofit     {\mbox{\textsc{RooFit}}\xspace}

\def\tell1  {TELL1\xspace}
\def\ukl1   {UKL1\xspace}

\newcommand{\eg}{\mbox{\itshape e.g.}\xspace}

\newcommand{\lhcborcid}[1]{\href{https://orcid.org/#1}{\hspace*{0.1em}\raisebox{-0.45ex}{\includegraphics[width=1em]{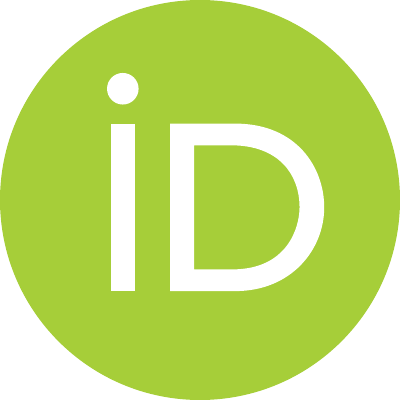}}}}

\usepackage{cite} 
\usepackage{mciteplus}

\def\histfactory     {\mbox{\textsc{HistFactory}}\xspace}
\def\RDst {\ensuremath{\mathcal{R}(\Dstar)}\xspace}
\def\RDstp {\ensuremath{\mathcal{R}(\Dstarp)}\xspace}
\def\RDstz {\ensuremath{\mathcal{R}(\Dstarz)}\xspace}

\def\RD {\ensuremath{\mathcal{R}(D)}\xspace}
\def\RDz {\ensuremath{\mathcal{R}(\Dz)}\xspace}

\def\RDb {\ensuremath{\mathcal{R}(D^{(*)})}\xspace}
\def\RJpsi {\ensuremath{\mathcal{R}(\jpsi)}\xspace}
\def\RLc {\ensuremath{\mathcal{R}(\Lc)}\xspace}

\def\RDeqinline {\ensuremath{\mathcal{R}(\D)\equiv{\mathcal{B}\left(\Bb\to\D\taum\neutb\right)}/
         {\mathcal{B}\left(\Bb\to\D\ellm\neulb\right)}}\xspace}
\def\RDsteqinline {\ensuremath{{\mathcal{R}(\Dstar)\equiv{\mathcal{B}\left(\Bb\to\Dstar\taum\neutb\right)}/{\mathcal{B}\left(\Bb\to\Dstar\ellm\neulb\right)}}}\xspace}
\def\RJpsieqinline {\ensuremath{{\mathcal{R}(\jpsi)\equiv{\mathcal{B}\left(\Bcm\to\jpsi\taum\neutb\right)}/{\mathcal{B}\left(\Bcm\to\jpsi\mun\neumb\right)}}}\xspace}
\def\RLceqinline {\ensuremath{\mathcal{R}(\Lc)\equiv{\mathcal{B}\left(\Lb\to\Lc\taum\neutb\right)}/{\mathcal{B}\left(\Lb\to\Lc\mun\neumb\right)}}\xspace}

\def\mmsq {\ensuremath{m^2_{\rm miss}}\xspace}
\def\El {\ensuremath{E^{*}_{\mu}}\xspace}
\def\qq {\ensuremath{q^{2}}\xspace}

\def\Done {\ensuremath{D^{}_1}\xspace}
\def\Donep {\ensuremath{D^{\prime}_1}\xspace}
\def\Dtwo {\ensuremath{D^{*}_2}\xspace}

\def\Dzero {\ensuremath{D^{*}_0}\xspace}

\def\dbtaunu {\decay{\Bb}{D^{(*)} \tau^{-} \neutb}}
\def\dblnu {\decay{\Bb}{D^{(*)} \ell^{-} \neulb}}
\def\dbmunu {\decay{\Bb}{D^{(*)} \mu^{-} \neumb}}
\def\dsttaunu {\decay{\Bzb}{D^{*+} \tau^{-} \neutb}}
\def\dstmunu {\decay{\Bzb}{D^{*+} \mu^{-} \neumb}}
\def\dstzmunu {\decay{\Bm}{D^{*0} \mu^{-} \neumb}}
\def\dstztaunu {\decay{\Bm}{D^{*0} \tau^{-} \neutb}}

\def\dstzLnu {\decay{\Bm}{D^{*0} l^{-} \neub_{l}}}

\def\dstarmunu {\decay{\Bb}{D^{*} \mu^{-} \neumb}}
\def\dstartaunu {\decay{\Bb}{D^{*} \tau^{-} \neutb}}

\def\dstlnu {\decay{\Bzb}{D^{*+} \ell^{-} \neub_{\ell}}}
\def\dstLnu {\decay{\Bzb}{D^{*+} l^{-} \neub_{l}}}

\def\dstarlnu {\decay{\Bb}{\Dstar \ell^{-} \neub_{\ell}}}

\def\dzmunu {\decay{\Bm}{D^{0} \mu^{-} \neumb}}

\def\dzLnu {\decay{\Bm}{D^{0} l^{-} \neub_{l}}}
\def\dlnu {\decay{\Bb}{D \ell^{-} \neulb}}
\def\dztaunu {\decay{\Bm}{D^{0} \tau^{-} \neutb}}

\def\dststmunu {\decay{\Bb}{D^{**} \mu^{-} \neumb}}
\def\dsststmunu {\decay{\Bsb}{D_{s}^{**} \mu^{-} \neumb}}

\def\dststlnu {\decay{\Bb}{D^{**} \ell^{-} \neulb}}

\def\dststtaunu {\decay{\Bb}{D^{**} \tau^{-} \neutb}}

\def\DbD {\decay{\B}{D^{(*)} D X}}
\def\DstDsTau {\decay{\Bb}{D^{*} D^{-}_{s} {(\to \taum \neutb)} X}}
\def\DbDsTau {\decay{\Bb}{D^{(*)} D^{-}_{s} {(\to \taum \neutb)} X}}

\def\DbHc {\ensuremath{\Bb\to \Dstarp H_{c}(\to\mun\neumb X^{\prime})X}\xspace}

\def\Dsstst {\ensuremath{D_s^{**}}\xspace}

\usepackage{longtable} 
\usepackage{subfigure}
\usepackage{overpic}
\usepackage{pbox}
\usepackage{pdfpages}
\usepackage{appendix}
\mciteErrorOnUnknownfalse
\usepackage{multirow}
\begin{document}

\renewcommand{\thefootnote}{\fnsymbol{footnote}}
\setcounter{footnote}{1}


\begin{titlepage}
\pagenumbering{roman}

\vspace*{-1.5cm}
\centerline{\large EUROPEAN ORGANIZATION FOR NUCLEAR RESEARCH (CERN)}
\vspace*{1.5cm}
\noindent
\begin{tabular*}{\linewidth}{lc@{\extracolsep{\fill}}r@{\extracolsep{0pt}}}
\ifthenelse{\boolean{pdflatex}}
{\vspace*{-1.5cm}\mbox{\!\!\!\includegraphics[width=.14\textwidth]{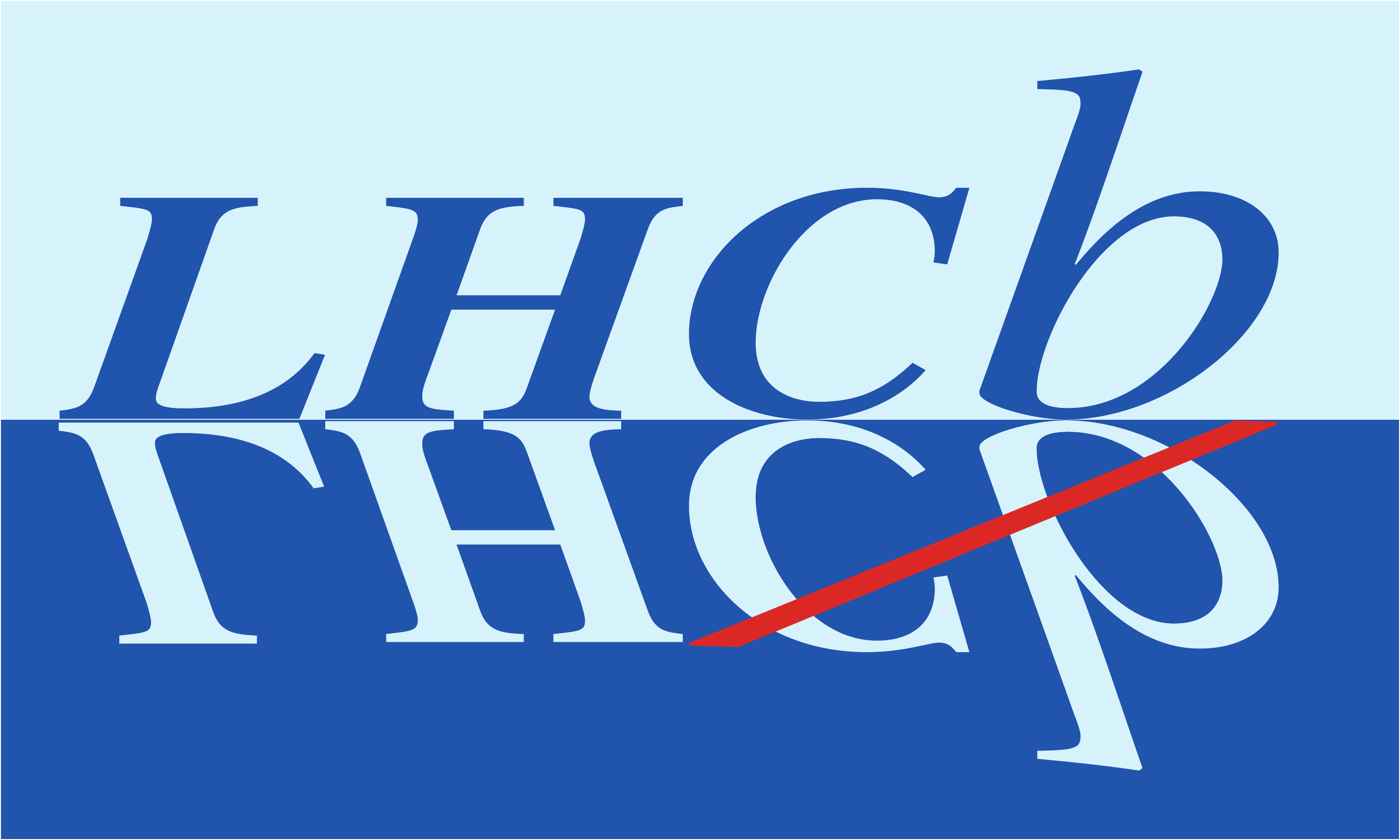}} & &}%
{\vspace*{-1.2cm}\mbox{\!\!\!\includegraphics[width=.12\textwidth]{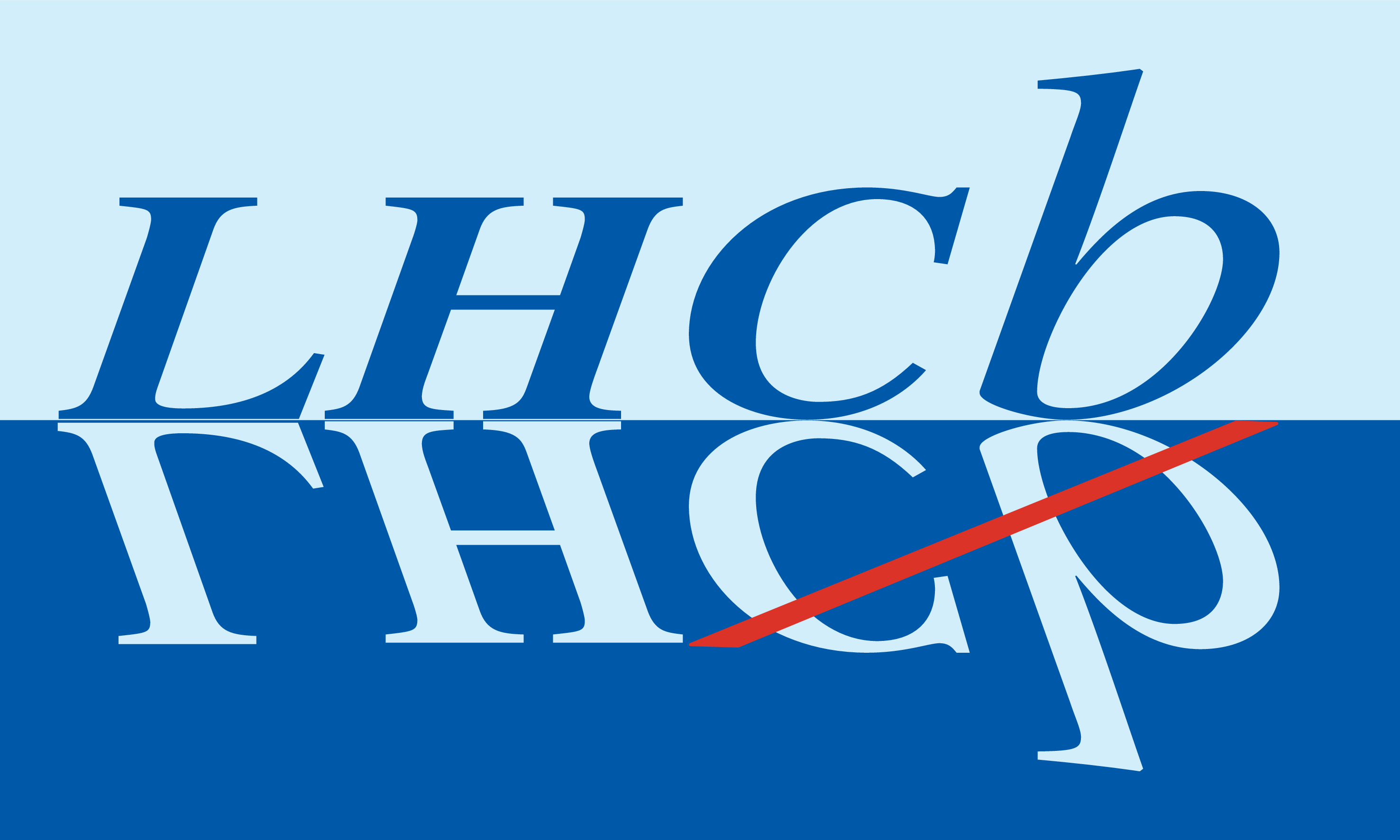}} & &}%
\\
 & & CERN-EP-2022-284 \\  
 & & LHCb-PAPER-2022-039 \\  
 & & September 19, 2023 \\ 
 & & \\
\end{tabular*}

\vspace*{2cm}

{\normalfont\bfseries\boldmath\huge
\begin{center}
  \papertitle 
\end{center}
}

\vspace*{2.0cm}

\begin{center}
\paperauthors\footnote{Authors are listed at the end of this Letter.}
\end{center}

\vspace{\fill}

\begin{abstract}
  \noindent
The ratios of branching fractions $\RDst\equiv\mathcal{B}(\dstartaunu)/\mathcal{B}(\dstarmunu)$ and $\RDz\equiv\mathcal{B}(\dztaunu)/\mathcal{B}(\dzmunu)$ are measured, assuming isospin symmetry, using a sample of proton-proton collision data corresponding to 3.0\invfb of integrated luminosity recorded by the LHCb experiment during 2011 and 2012.  The tau lepton is identified in the decay mode 
 $\taum\to\mun\neut\neumb$.  The measured values are $\RDst=0.281\pm0.018\pm0.024$ and $\RDz=0.441\pm0.060\pm0.066$, where the first uncertainty is statistical and the second is systematic. The correlation between these measurements is $\rho=-0.43$. Results are consistent with the current average of these quantities and are at a combined 1.9 standard deviations from the predictions based on lepton flavor universality in the Standard Model.

\end{abstract}

\vspace*{2.0cm}

\begin{center}
Published in Phys.~Rev.~Lett.~131~(2023)~111802
\end{center}

\vspace{\fill}

{\footnotesize 
\centerline{\copyright~\papercopyright. \href{\paperlicenceurl}{\paperlicence}.}}
\vspace*{2mm}

\end{titlepage}


\newpage
\setcounter{page}{2}
\mbox{~}


\renewcommand{\thefootnote}{\arabic{footnote}}
\setcounter{footnote}{0}

\cleardoublepage


\pagestyle{plain} 
\setcounter{page}{1}
\pagenumbering{arabic}


Semileptonic \bquark-hadron decays provide a powerful laboratory for testing the equality of the couplings of the three charged leptons to the gauge bosons, a fundamental characteristic of the Standard Model (SM), known as lepton flavor universality (LFU).
Measurements of the LFU-sensitive ratios of branching fractions \RDsteqinline, where \Dstar indicates either \Dstarp or \Dstarz and $\ell$ indicates a light lepton,\footnote{The inclusion of charge-conjugate processes is implied throughout this Letter.} \RDeqinline
\cite{BaBar:2013mob,LHCb-PAPER-2015-025,Belle:2015qfa,Belle:2017ilt,LHCb-PAPER-2017-017,LHCb-PAPER-2017-027,Belle:2019rba}, where \D indicates \Dp or \Dz,
and \RJpsieqinline \cite{LHCb-PAPER-2017-035}
show an excess of semitauonic decays over the SM predictions,whereas a measurement of \RLceqinline \cite{LHCb-PAPER-2021-044} is found to be consistent with the SM. A summary of predictions of \RD and \RDst in particular is presented in Table \ref{tab:SM}.  

\begin{table}[bh]
\caption{Summary of calculations of \RD and \RDst as compiled by the HFLAV collaboration \cite{HFLAV:2022pwe}. For consistency with HFLAV this Letter uses the same average value $\RD = 0.298\pm 0.004$, $\RDst=0.254\pm 0.005$ when making comparisons with the Standard Model.} 
\label{tab:SM}
\begin{center}
\begin{tabular}{c c c}
\hline\hline
\RD & \RDst & Ref.\\
\hline
$0.299\pm0.003$ & -  & \cite{Bigi:2016mdz}\\
- & $0.254\,{ }^{+0.007}_{-0.006}$  & \cite{Gambino:2019sif} \\
$0.298 \pm 0.003$ & $0.247 \pm 0.006$  & \cite{Bordone:2019vic} \\
$0.299\pm0.003$ & $0.257 \pm 0.003$  & \cite{Bernlochner:2017jka} \\
$0.299 \pm 0.004$ & $0.257 \pm 0.005$  & \cite{Jaiswal:2017rve} \\
- & $0.253\pm0.005$  & \cite{BaBar:2019vpl} \\
$0.296 \pm 0.008$ & -  & \cite{Martinelli:2021onb}\\
- & $0.260 \pm 0.008$  & \cite{Bigi:2017jbd} \\
- & $0.265 \pm 0.013$  & \cite{FermilabLattice:2021cdg}
\end{tabular}
\end{center}
\end{table}

The LHCb collaboration has previously reported on LFU studies in the $\bquark\to \cquark$ semileptonic 
decays using the data recorded during 2011--2012: two measurements of \RDstp using the purely 
leptonic tau decays $\taum\to\mun\neut\neumb$ \cite{LHCb-PAPER-2015-025} and the three-pion decay 
channel $\taum\to\pim\pip\pim(\piz)\neut$ \cite{LHCb-PAPER-2017-017,LHCb-PAPER-2017-027}, as well as measurements 
of the observables \RJpsi in the leptonic channel \cite{LHCb-PAPER-2017-035} and \RLc in the three-pion channel \cite{LHCb-PAPER-2021-044}. 
This Letter presents the first simultaneous measurement of \RDst and \RDz in hadron collisions, and supersedes the result of Ref. \cite{LHCb-PAPER-2015-025}.
The data correspond to integrated 
luminosities of 1.0\invfb and 2.0\invfb, collected by
the LHCb detector in proton-proton collisions with
center-of-mass 
energies of 7 TeV and 8 TeV, respectively. Owing to the different spin structures of the 
\dstarlnu and \dlnu decays, the combined result provides significantly improved sensitivity to the structure of 
possible LFU-breaking processes originating from physics beyond the SM, such as the effects of an extended Higgs mechanism or leptoquarks (see, \eg, the recent review in Ref.~\cite{Bernlochner:2021vlv}).

This study utilizes the purely leptonic tau decay  $\taum\to\mun\neut\neumb$ for the reconstruction 
of the semitauonic \dbtaunu decays, where $D^{(*)}$ stands for a \Dz,  a \Dstarp or a \Dstarz meson.
These decays,  hereafter denoted as the signal channels,  as well as \dbmunu decays, 
which serve as the normalization for the determination of the \RDz and \RDst observables, are identified 
using the visible final states $\Dz\mun$ and \Dstarp\mun. 
Both signal and normalization channels are selected by a common reconstruction procedure, which selects events containing 
a muon candidate and a $\Dz\to\Km\pip$ candidate with the expected flavor correlation, $\Dz\mun$, from 
$\bquark\to \cquark$ semileptonic decays. The sample is divided into $\Dz\mun$ and \Dstarp\mun samples according to whether the combination of the \Dz with any track in the event forms a \Dstarp candidate with a mass difference $\Delta m < 160 \mevcc$, where $\Delta m$ is the difference between the \Dstarp and \Dz candidate masses. 

The $\Dz\mun$ sample contains 
contributions from \dztaunu and \dzmunu decays as well as 
 contributions from partially reconstructed  \dstzmunu, \dstmunu, \dstztaunu, and \dsttaunu decays through the decay
 chains ${\Dstarp\to\Dz\pip}$, ${\Dstarz\to\Dz\gamma}$ and ${\Dstarz\to\Dz\piz}$, where the photon or pion is not reconstructed. 
The $\Dstarp\mun$ candidate sample, 
which contains \dsttaunu and \dstmunu and no substantial contribution from the other signal or normalization decays, was the basis of the first  measurement of \RDst by the LHCb collaboration \cite{LHCb-PAPER-2015-025}. The simultaneous analysis of the two samples helps to constrain the common parameters of the fit models that are applied to the data, reducing the correlation between the measured values of \RDz and \RDst.

In addition to the signal and the normalization channels, the selected samples contain contributions from several background processes, which include: 
partially reconstructed \B decays, such as semileptonic decays with an excited charmed meson and hadronic \B decays into two charmed mesons, with one of them decaying (semi)leptonically; cases where the muon candidate originates from the misidentification of other charged particles; and combinations of unrelated particles from different decay chains.  
The kinematic and the topological properties of the various components are exploited to
suppress background contributions. The relative contributions of the processes
present in the data samples are determined by fitting to the data 
a model composed of multidimensional template distributions derived from
control samples in data or from simulation validated against data.

\label{sec:Detector}

The \lhcb detector, described in detail in Refs.~\cite{LHCb-DP-2008-001,LHCb-DP-2014-002}, is a single-arm forward
spectrometer covering the \mbox{pseudorapidity} range $2<\eta <5$,
designed for the study of particles containing \bquark or \cquark
quarks. 
The online event selection is performed by a trigger~\cite{LHCb-DP-2012-004}, 
which consists of a hardware stage, based on information from the calorimeter and muon
systems, followed by a software stage, which applies a full event
reconstruction.
Simulation produced by software packages described in Refs.~\cite{Sjostrand:2006za,Sjostrand:2007gs,Lange:2001uf, Golonka:2005pn,Allison:2006ve, Agostinelli:2002hh}
is used to model the physics processes and the effects of both the detector acceptance and the selection criteria.

The events are required to pass the hardware trigger independently of the 
muon candidate, as the requirement in the hardware trigger on the 
component of the muon momentum transverse to the beam, $\pt(\mu)$, 
would significantly reduce the selection efficiency of the semitauonic 
decays. Therefore, the events must pass the hardware trigger either 
because the decay products of the $D^0\to K^-\pi^+$ candidate satisfy 
the hadron trigger requirements or because unrelated high-\pt 
particles in the event satisfy any of the hardware trigger requirements. 
In the software trigger, the events
are required to meet criteria designed to accept $\Dz\to\Km\pip$ candidates 
with $\pt> 2\gevc$.  Quality requirements are applied to the tracks of the charged particles that originate from a candidate \Dz decay: their momenta must exceed 5\gevc and at least 
one track must have 
 $\pt> 1.5 \gevc$.  The momentum vector of the \Dz candidate must align approximately with the displacement from one of the primary vertices (PV) in 
 the event and the reconstructed mass must be consistent with the known \Dz mass \cite{PDG2022}. 
 
In the offline reconstruction, \Km and \pip candidates from the \Dz decay are required to satisfy loose particle identification requirements, and the decay vertex is required to be significantly displaced 
from all PVs. The invariant mass of the \Dz candidate is required to be consistent with the \Dz mass within three times the resolution, as determined by a fit to data. 
The muon candidate is required to be 
consistent with a muon signature in the detector, to have momentum  $3 <p< 100 \gevc$,  
to be significantly separated from any PV, and to form a good vertex with the $D^0$ candidate. 
To reduce further the background from hadrons misidentified as muons (``misID background''), muon likelihood-ratio identification criteria used previously \cite{LHCb-PAPER-2015-025} are supplemented with a dedicated multivariate selector, trained on information from multiple subdetectors, constrained to provide uniform efficiency in muon momentum and $\pt$ using the uBoost method \cite{Stevens:2013dya}. 
The $\Dz\mun$ combinations are required to have an invariant
mass less than 5280\mevcc and their momentum vector is required to align approximately with the displacement 
vector from the associated PV to the \Dz\mun vertex, which removes random combinations  while preserving a large 
fraction of semileptonic decays. In addition to the signal candidates, independent samples of 
``wrong sign" candidates, $\Dz\mup$, $\Dstarp\mup$ and $\Dz\pim\mun$, are 
selected for estimating the combinatorial background. The first two represent random combinations of 
$D^{(*)}$ candidates with muons from unrelated decays, and the latter is used to model the contribution 
of misreconstructed $\Dstarp$ decays. The background from misreconstructed $\Dz\to\Km\pip$ candidates is negligible. 
Mass regions ${5.28 < m(\Dz\mun) < 10\gevcc}$ and ${\Delta m < 160\mevcc}$ are included in all samples 
to study the combinatorial backgrounds. 

To suppress the contributions from partially reconstructed \B decays, 
 the signal candidates are required to be isolated from additional tracks in the event. The isolation algorithm is described in Ref.~\cite{LHCb-PAPER-2015-025}. 
Except for the muon identification procedure, the selection criteria for the $D^{*+}\mu^-$ sample are unchanged from those used in Ref.~\cite{LHCb-PAPER-2015-025}.

Kinematic variables in the \B candidate rest frame, approximated from the laboratory quantities by taking the \B boost along the beam axis to be equal to that of the visible candidate \cite{LHCb-PAPER-2015-025}, are used to characterize and discriminate between the various processes.
These variables are: 
the muon energy in the \B rest frame, $E_{\mu}^*$; the missing mass squared, defined as
 $m^2_{\textrm{miss}}=(p_{\B} \!-\! p_{D^{(*)}} \!-\! p_{\mu})^2$; and the squared four-momentum transfer to the lepton system, $q^2=(p_{\B} - p_{D^{(*)}})^2$, where $p_{\B}$, $p_{D^{(*)}}$ and $p_{\mu}$ are the four-momenta of 
 the \B meson, the \Dz (or \Dstarp) meson and the muon.

Simulated events are used to derive the distributions of the \B candidate rest-frame kinematic variables 
for the signal and the normalization channels as well as background from other partially reconstructed \bquark-hadron decays.  The sum of 
these multidimensional template distributions, which depends on shape parameters and the relative 
yields of the contributing processes, forms the fit model applied to the data.
Two independent fit model implementations and associated frameworks using common form factor and correction parameterizations have 
been developed and applied to the data allowing for cross-checks on nearly all aspects of the analysis. 

The simulated samples of $\Bb\to\Dstar\mun\neumb$ and $\Bb\to\Dstar\taum\neutb$ decays are weighted to the
Boyd, Grinstein, and Lebed (BGL) form-factor parameterization~\cite{Boyd:1997kz} using values presented in Refs.~\cite{Bigi:2016mdz,Bigi:2017jbd,Jaiswal:2017rve} as a starting point. 
For the decays \dzmunu and \dztaunu,  form factors are described using the parameterization of Refs.~\cite{Na:2015kha, BCL}. 
For the results presented here, the BGL form factor expansion coefficients up to order $z^2$ are allowed to vary in the fit with loose likelihood constraints only for those describing the helicity-suppressed form factors.

The backgrounds from semileptonic \Bb decays to the lowest-lying excited charm states $\Done(2420)$, 
$\Dtwo(2460)$, $\Donep(2430)$, and $\Dzero(2300)$ (collectively referred to as $\D^{**}$) are weighted to the form-factor parameterization presented in
Ref.~\cite{Bernlochner:2016bci}.
These form-factor parameters, which are allowed to vary without constraint in the fit, are constrained by 
control regions in the data, described below. Background from $\Bsb$ decays to the states $D_{s1}(2536)^{+}$ 
and $D^{*}_{s2}(2573)^+$, (together denoted as \Dsstst) which subsequently decay as 
$\D_s^{**+}\to D^{*+}K^0$ or $\D_s^{**+}\to D^{0}K^{+}$, are modeled using the same 
form-factor parameterization, with values unconstrained and independent of those for the $D^{**}$ states. 
Backgrounds from semileptonic $B$ decays to heavier excited charm states $D^{**}_{\mathrm{heavy}}$ decaying as $D^{**}_{\mathrm{heavy}}\to D^{(*)}\pi\pi$ are modeled using simulated samples containing a mix of final states generated with the ISGW2~\cite{ISGW2} form factors. As the composition and decay properties of this background are not well understood, an ad-hoc weight is applied as a linear function of the true \qq with independent slopes for decays to \Dstarp, \Dstarz, and \Dz, which vary in the fit and are constrained by control regions in the data. 

For the background from \B decays into two charmed mesons, simulated samples of \Bzb and \Bm decays $\B \to \Dstarp H_c X$ 
and $\B \to \Dz H_c X$ with a mix of final states are used, where $H_c$ is a
charm meson that decays (semi)leptonically, yielding a correct-sign secondary muon to combine with 
the $\Dz$ or $\Dstarp$, and $X$ is any combination of light hadrons (\eg, a $K$ or $K^{*}$ meson). 
The multibody decays are simulated uniformly in phase space. Control samples (discussed below) are included to constrain corrections applied to the decay distribution. The corrections involve weights given by linear and quadratic functions of the invariant mass of the two primary charm hadrons as well as variations of the size of the contribution 
of modes with $m(X)> 680\mevcc$, where the additional particles $X$ are mostly $K^*$ resonances. A separate sample is used to model the contribution from tertiary muons from $H_c = D_s^{(*(*))-}$ 
with the leptonic decay $D_s^- \to \taum \neutb$.

To model the contribution of misID background, a control sample of \Dz or \Dstarp 
candidates paired with a single track is used, where the combinations pass all the analysis
selection criteria but the single track 
has no associated
segment in the muon system. The two fit models employ two different techniques to
produce a model of the misidentified backgrounds by weighting this control sample.
Both techniques produce per-track weights using particle identification classification information on the extra track, \pion, \kaon, \proton, or \electron, or unidentified, combined with particle identification efficiencies from large calibration samples. 
Details are given in the Supplemental Material \cite{ref:SuppMat}.
Both techniques are independently validated by fitting
data samples in which the muon candidate passes initial muon identification criteria, but
fails to pass the custom multivariate muon identification developed for this analysis.

Combinatorial backgrounds are classified based on whether or not a genuine ${\Dstarp \to \Dz\pip}$ decay is present. 
Wrong-sign $\Dz\pim\mun$ 
combinations are used 
to determine the component with misreconstructed \Dstarp candidates. The size of this contribution is constrained by a fit to the $\Delta m$ distribution of $\Dstarp\mun$ candidates.
The contribution from correctly reconstructed \Dstarp candidates combined with $\mun$ from unrelated $b$-hadron decays is determined from wrong-sign $D^{(*)}\mup$ 
combinations. 
In both cases, the contributions of misidentified muons are subtracted when generating the kinematic distributions for the fit. The mass region $5.28 < m(D^{(*)}\mu^{\mp}) < 10 \gevcc$ excludes genuine \B decays, and is used to validate the agreement between the kinematic distributions for wrong-sign and correct-sign combinatorial background candidates.

Extensive studies are performed to account for differences between the data and the simulation.
An initial set of corrections to the $b$-hadron production distributions is applied based 
on weights derived from the comparison of a sample of reconstructed $\Bp \to \jpsi \Kp$ decays in data and simulation. A sequence of additional corrections, as a function of kinematic and topological variables, is then applied to the simulation using a control sample of $D^0\mun$ candidates with $\mmsq < 0.4 \gevgevcccc$, which is dominated by the \dbmunu decay. 
Further corrections are applied using the corresponding region of the $\Dstarp\mun$ sample to correct residual differences between data and simulation in the reconstruction of the low-momentum \pip in the $\Dstarp\to\Dz\pip$ decay. The combined correction results in an RMS bin-by-bin change to the fit templates on the order of 3\%, and improves the modelling of both the acceptance and the resolution for the fit variables (and therefore bin migration effects), including correlations. This procedure is, in principle, iterative, but converges after a single round of corrections, and residual differences are covered in the systematic uncertainties discussed below.

In order to constrain the modeling of the various backgrounds presented above, 
several control regions enhanced in the background contributions are selected in both the \Dz\mun and \Dstarp\mun data based on the output of the isolation algorithm. Requiring the presence 
of a charged kaon candidate among the particles accompanying the $D^{(0,*+)}\mun$ candidate 
results in a sample with an enhanced fraction of \B decays into two charmed 
mesons. Samples enriched in semileptonic \B decays to $D^{**}$ are selected by 
requiring the presence of exactly one additional pion candidate in the vicinity of the 
$D^{(0,*+)}\mun$ candidate with the correct relative charge for the $D^{**}\to D^{(*)}\pi$ 
decay. Requiring exactly two accompanying pion candidates of opposite charge provides a 
sample with enhanced fraction of decays to $D^{**}_{\rm heavy}$ mesons. The three isolation output selections above, selected in both $\Dz\mun$ and $\Dstarp\mun$ samples, constitute six distinct control regions. Including the isolated $\Dz\mun$ and $\Dstarp\mun$ signal region, eight regions are selected in total for the fit.

The binned distributions of \mmsq ($[-2,10.6]$\gevgevcccc, 43~bins), \El ($[100,2650]$\mev, 34~bins), and \qq ($[-0.4,12.6]$\gevgevcccc, 4~bins) for reconstructed $\Dz\mun$ and $\Dstarp\mun$ 
candidates in data are fit using a binned extended maximum-likelihood method with three-dimensional 
templates representing the signal and normalization channels and the background sources. The model parameters 
extracted from the data include the yields of each contributing process: signals, normalizations, \dststlnu (with  
a Gaussian-constrained fraction of \dststtaunu), $\Bb\to D^{**}_{\mathrm{heavy}}\mun\neumb$, \dsststmunu, \DbHc (with  
a Gaussian-constrained fraction of \DbDsTau), misID background, and combinatorial backgrounds. 
The form-factor parameters for signal, normalization, and $\D_{(s)}^{**}$ backgrounds are allowed 
to vary in the fit, 
as is the level of momentum smearing applied to the misID component to account for kaon or pion decays to muons. The same fit model is applied to all selected regions with appropriately selected templates, with form-factor parameters and shape correction parameters shared between regions, and yield parameters allowed to vary independently by region.
Statistical uncertainties in the templates are folded into the likelihood via the Beeston-Barlow `lite' prescription \cite{Barlow:1993hy}. Projections of the fit in each control region are shown in the Supplemental Material~\cite{ref:SuppMat}.

 Two
approaches are used to incorporate the information from the
control regions. For the result presented here, all eight regions are fit simultaneously using a custom likelihood implementation in the ROOT~\cite{Brun:1997pa} software package to
extract \RDz and \RDst including all correlations. In the alternative fit, built using the \roofit~\cite{Verkerke:2003ir} and \histfactory~\cite{Cranmer:1456844} frameworks, the six
control regions are fit simultaneously first
to obtain corrections to the  
most signal-like backgrounds in signal-depleted regions. The two signal samples are then fit with shapes
fixed (or likelihood-constrained in the case of the \dststmunu and \dsststmunu form-factor parameters) according 
to the result of the control fit. The two fitters have been extensively cross-validated and give consistent
results within an expected statistical spread determined using common pseudodatasets.
As the two results are compatible, only the results of the former fit are presented in this Letter.

\begin{figure*}[tbp]
    \begin{center}
    \includegraphics[trim=0cm 0cm 0.1cm 0cm, clip=True,width=0.81\textwidth]{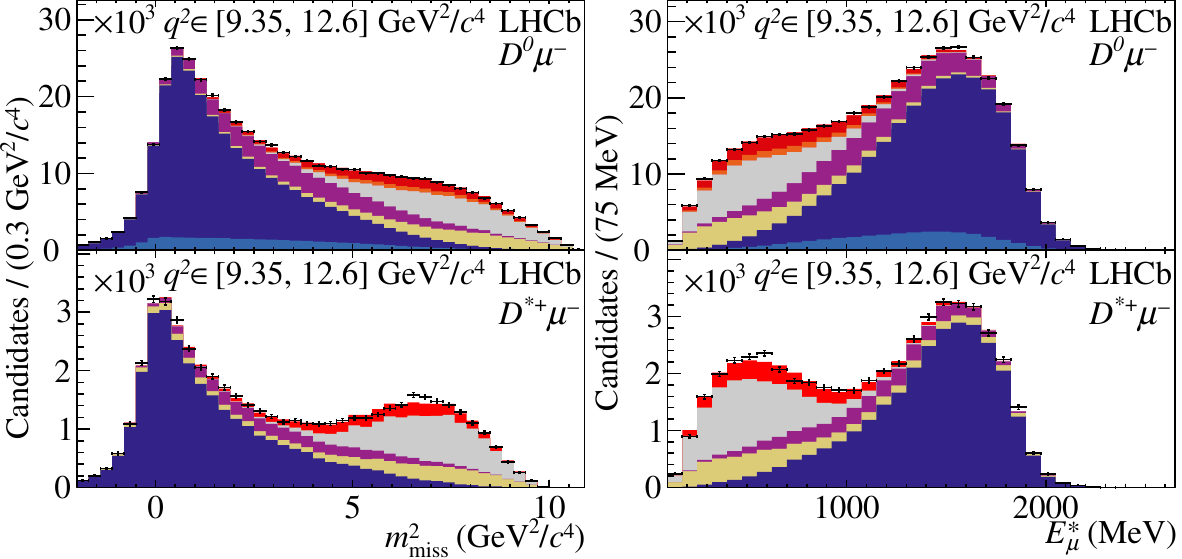}
    \raisebox{1.1cm}{\includegraphics[trim=0.3cm 0cm 0.12cm 0cm, clip=True, width=0.18\textwidth]{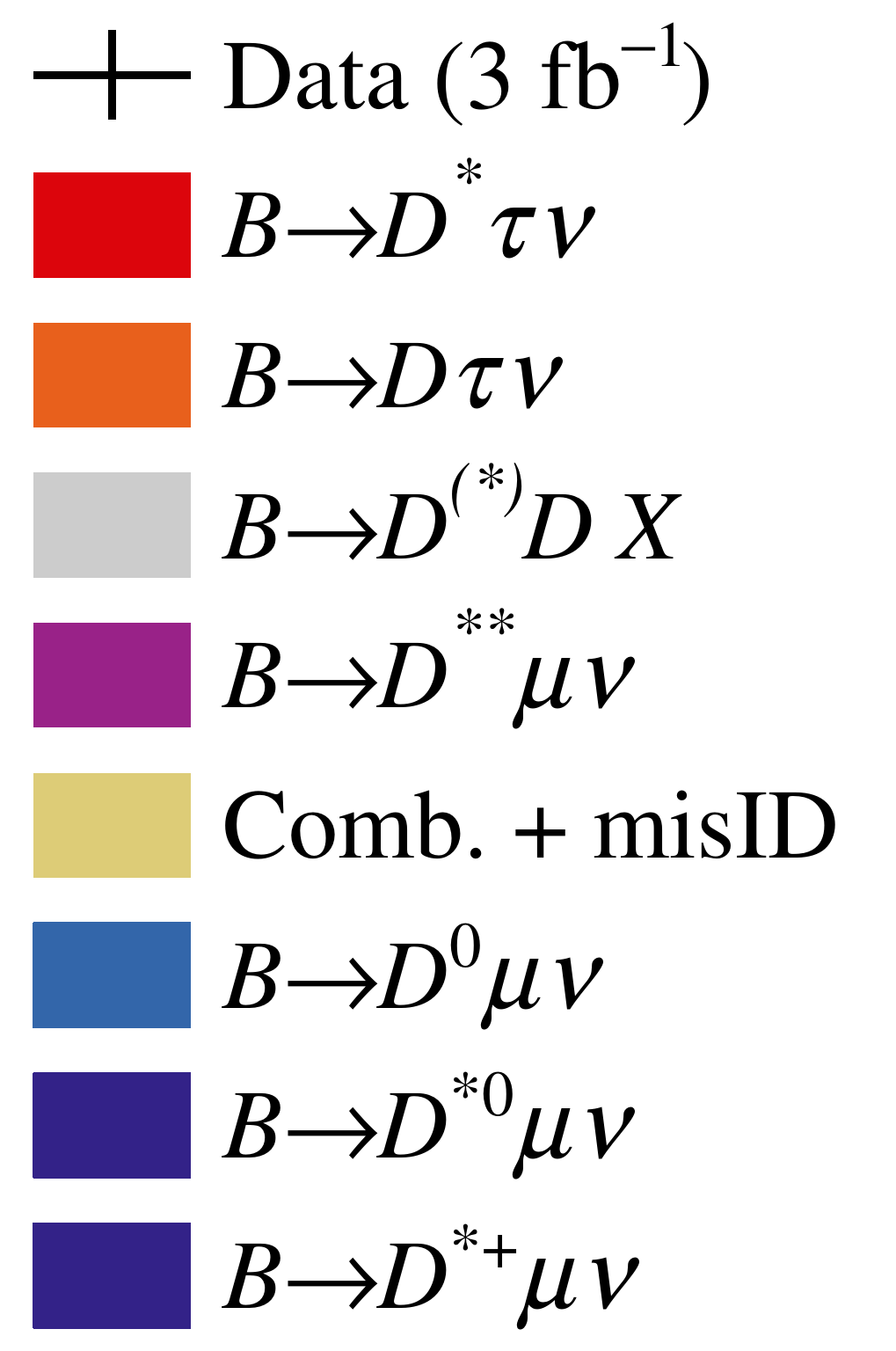}}
    \end{center}
    \caption{Distributions of (left) $m^2_{\rm miss}$ and (right) \El in the highest \qq bin (above 9.35\gevgevcccc) of the (top) \Dz\mun  and (bottom) \Dstarp\mun signal data, overlaid with projections of the fit model.}
    \label{fig:theFit}
    \end{figure*}
    The results of the fit to the isolated (signal) samples are shown in Fig.~\ref{fig:theFit}. The complete set of projections for all \qq bins can be found in the Supplemental Material~\cite{ref:SuppMat}. The ratios of branching fractions are determined to be
    $\RDz=0.441\pm0.060$, $\RDst=0.281\pm0.018$, with a correlation $\rho=-0.49$, where the statistical uncertainties are evaluated with all nuisance parameters related to template shape uncertainties fixed to their respective best-fit values. 
        The obtained yields of normalization (signal) are $324\,000$ ($12\,000$) \dstLnu decays in the $\Dstarp\mun$ signal sample, and $354\,000$ ($20\,000$) \dzLnu decays, $958\,000$ ($34\,000$) \dstzLnu decays, and
    $44\,000$ ($1\,700$) \dstLnu decays in the $\Dz\mun$ sample, where $l=\mu$ ($\tau$).
    The  \dstmunu yield in the $\Dstarp\mun$ sample is consistent with the previous measurement \cite{LHCb-PAPER-2015-025} after accounting for the efficiency of the stricter
    muon identification criteria used here.

\begin{table*}[bt]
    \caption{Absolute uncertainties in the extraction of \RDz and \RDst. The model uncertainties are divided into those included directly in the fit likelihood and those determined via supplemental studies.}
      \centering
        \begin{tabular}{ l  c  c  c }
          \hline\hline
          \textbf{Internal fit uncertainties} & $\mathbf{\sigma}_{\RDst}$\scalebox{0.85}{$(\times \!10^{-2})$} & $\mathbf{\sigma}_{\RDz}$\scalebox{0.85}{$(\times \!10^{-2})$} & \textbf{Correlation} \\
          \hline
          Statistical uncertainty & 1.8 & 6.0 & $-0.49$ \\
          Simulated sample size & 1.5 & 4.5 &\\
          \DbD template shape     & 0.8 &  3.2 &\\
          \dblnu form-factors     & 0.7 & 2.1 &\\
          \dststmunu form-factors  & 0.8 & 1.2 &\\
          \BR(\DstDsTau)           &  0.3 & 1.2 &\\
          MisID template       & 0.1 & 0.8 &\\
          \BR(\dststtaunu)         & 0.5 & 0.5 &\\
          Combinatorial       & $<0.1$ & 0.1 &\\
          Resolution       & $<0.1$ & 0.1 &\\
          \hline
          \textbf{Additional model uncertainty}& $\mathbf{\sigma}_{\RDst}$\scalebox{0.85}{$(\times \!10^{-2})$} & $\mathbf{\sigma}_{\RDz}$\scalebox{0.85}{$(\times \!10^{-2})$} & \\
          \hline
          \DbD model uncertainty       & 0.6 & 0.7 &\\
          \dsststmunu model uncertainty       & 0.6 & 2.4 &\\
          Baryonic backgrounds & 0.7 & 1.2 &\\
          Coulomb correction to $\RDstp/\RDstz$     & 0.2 & 0.3 &\\
          Data/simulation corrections       & 0.4 & 0.8 &\\
          MisID template unfolding & 0.7 & 1.2 &\\
          \hline
          \textbf{Normalization uncertainties} & $\mathbf{\sigma}_{\RDst}$\scalebox{0.85}{$(\times \!10^{-2})$} & $\mathbf{\sigma}_{\RDz}$\scalebox{0.85}{$(\times \!10^{-2})$} & \\\hline
          Data/simulation corrections       & 0.4$\times \RDst$ & 0.6$\times \RDz$ &\\
          $\taum\to\mun\nu\neub$ branching fraction & 0.2$\times \RDst$ & 0.2$\times \RDz$ & \\
      
          \hline
            \textbf{Total systematic uncertainty} & 2.4 & 6.6 & $-0.39$\\
          \textbf{Total uncertainty} & 3.0 & 8.9 & $-0.43$\\
          \hline
          \hline
        \end{tabular}
      
      \label{tab:systematics}
    \end{table*}

    Uncertainties in the measurements of \RDz and \RDst are summarized in Table~\ref{tab:systematics}. The uncertainty in extracting \RDb from the fit (model uncertainty) is dominated by the 
    statistical uncertainty of the simulated samples; this contribution is estimated via the reduction in the fit uncertainty when the template statistical uncertainty is not considered in the likelihood.  
    Form-factor parameters are included in the likelihood as nuisance parameters,  
    hence the associated systematic uncertainties are contained in the total uncertainties of \RDz and \RDst determined with all nuisance parameters allowed to vary.
    To determine the contribution of the form-factor uncertainty, the fit is repeated with form-factor parameters fixed to their best-fit values, and the reduction in uncertainty compared with the configuration with varying nuisance parameters is used to determine the contribution from the form-factor uncertainties.
    The systematic uncertainty from empirical corrections to the kinematic distributions of $\Bb \to D^{**}(\to D^{(*)}\pi\pi)\mun\neumb$ and $\Bb \to D^{(*)} H_{c}(\to\mu\neum X^{\prime})X$ backgrounds is computed in the same way. 
    
    The contribution of ${\B \to D^{**}_{(s)}\taum\neutb}$ decays relative to ${\B \to D^{**}_{(s)}\mun\neumb}$ is likelihood-constrained to an expectation of $8\%$ taken from Ref.~\cite{Bernlochner:2016bci}, with a relative uncertainty of 30\% assigned to cover both the inclusion of different $D^{**}_{(s)}\taum\neutb$ states, and the possibility of LFU violation in these decay modes.
    Similarly, the contribution of ${\Bb \to D^{(*)} D^{-}_s (\to \taum \neutb)X}$ decays is likelihood-constrained 
    using known branching fractions \cite{PDG2022} with a $30\%$ uncertainty. The systematic uncertainty is again given by the effect of allowing these to vary within the loose $30\%$ constraint versus fixing them to best-fit.

The uncertainty in the shape of the misID background due to the limited statistics of control data is determined by allowing the hadron to muon misidentification efficiency vary as a function of lab momentum within the uncertainties from control data and computing the increase in uncertainty in \RDb.

 The expected yield of $D^{(*)}\mun$ candidates compared to $D^{(*)}\mup$ candidates (used to model the 
    combinatorial background) varies as a function of $m(D^{(*)}\mu^{\mp})$. The size of this effect is estimated in the $5.28 < m(D^{(*)}\mu^{\mp}) < 10 \gevcc$ region and the uncertainty is propagated as a systematic uncertainty in \RDb.
    
    The choice of corrections applied to simulated $\B \to D^{(*)} H_c X$ decays is not unique, and so the fit is repeated for an ensemble of possible alternative choices. The root mean square of this ensemble is taken as a systematic uncertainty. 
    
    A small discrepancy in the fit quality is observed in a region of the control samples dominated by cross-feed from \dsststmunu decays. To assess the maximum size of the effect from this mismodelling, a deformation suppressing the low-\qq, low-\El region of this template to better match the data is applied and the effect on the signal yield from this change is evaluated. 
    
    The default fit model does not include $\Lb \to \Dz p \mun \neumb$ or $\Lb \to \Dstarp n \mun \neumb$ decays. To assess the effect of their exclusion, a fit is performed to a control sample requiring a proton candidate among the particles accompanying the $\Dz \mun$ candidate. The existing \dststmunu simulated samples are reused with different parameter values as proxy for the \Lb decays and are able to reproduce the kinematic distributions observed in the data. The fit for \RDb is repeated with these components included using one of two possible auxilliary fits to constrain the size of the contribution, and the larger of the two possible shifts of \RDb is assigned as the systematic uncertainty.

    The systematic uncertainty due to the absence of the Coulomb interaction in the \photos package~\cite{Golonka:2005pn} is evaluated by weighting the \dstlnu simulation by the Coulomb factor given in Ref.~\cite{Cali:2019nwp}.
    It is found that the only significant effect on these results is due to a 1\% shift of the expected isospin relationship between \RDstp and \RDstz, which induces a small shift in \RDz and \RDst.
    
    To assess the uncertainty from residual disagreements between data and simulation, a second iteration of the weighting procedure described above is performed using several possible variations of the scheme. Half the largest difference in \RDb is taken as a systematic uncertainty.

The systematic uncertainty from the unfolded kinematic shapes of the misID background is taken to be half the difference from using the two misID determination methods described above.
    
    Uncertainties in converting the fitted ratio of signal and normalization yields into \RDb (normalization uncertainties) primarily come from the uncertainty in the effect of the corrections to simulation and are evaluated similarly. 
    The uncertainty in the current world average value of $\mathcal{B}(\taum\to\mun\neumb\neut)$ also contributes a small normalization uncertainty.

In conclusion, the branching fraction ratios $\mathcal{B}(\dstartaunu)/\mathcal{B}(\dstarmunu)$ and $\mathcal{B}(\dztaunu)/\mathcal{B}(\dzmunu)$ 
are measured to be 
\begin{align*}
\RDst &= 0.281\pm0.018\pm0.024 \\
\RDz &= 0.441\pm0.060\pm0.066 \\
\rho  &= -0.43, 
\end{align*} 
with $\rho$ the correlation, and where the first uncertainty is statistical and the second is systematic. This is the first measurement of the ratio $\RDz$ at a hadron collider. These results are consistent at less than one standard deviation with the current average of these quantities and stand at about two standard deviations from the predictions based on lepton flavor universality in the Standard Model.

\section*{Acknowledgements}
%
%
\noindent We express our gratitude to our colleagues in the CERN
accelerator departments for the excellent performance of the LHC. We
thank the technical and administrative staff at the LHCb
institutes.
We acknowledge support from CERN and from the national agencies:
CAPES, CNPq, FAPERJ and FINEP (Brazil); 
MOST and NSFC (China); 
CNRS/IN2P3 (France); 
BMBF, DFG and MPG (Germany); 
INFN (Italy); 
NWO (Netherlands); 
MNiSW and NCN (Poland); 
MCID/IFA (Romania); 
MICINN (Spain); 
SNSF and SER (Switzerland); 
NASU (Ukraine); 
STFC (United Kingdom); 
DOE NP and NSF (USA).
We acknowledge the computing resources that are provided by CERN, IN2P3
(France), KIT and DESY (Germany), INFN (Italy), SURF (Netherlands),
PIC (Spain), GridPP (United Kingdom), 
CSCS (Switzerland), IFIN-HH (Romania), CBPF (Brazil),
Polish WLCG  (Poland) and NERSC (USA).
We are indebted to the communities behind the multiple open-source
software packages on which we depend.
Individual groups or members have received support from
ARC and ARDC (Australia);
Minciencias (Colombia);
AvH Foundation (Germany);
EPLANET, Marie Sk\l{}odowska-Curie Actions and ERC (European Union);
A*MIDEX, ANR, IPhU and Labex P2IO, and R\'{e}gion Auvergne-Rh\^{o}ne-Alpes (France);
Key Research Program of Frontier Sciences of CAS, CAS PIFI, CAS CCEPP, 
Fundamental Research Funds for the Central Universities, 
and Sci. \& Tech. Program of Guangzhou (China);
GVA, XuntaGal, GENCAT and Prog.~Atracci\'on Talento, CM (Spain);
SRC (Sweden);
the Leverhulme Trust, the Royal Society
 and UKRI (United Kingdom).

\addcontentsline{toc}{section}{References}
\bibliographystyle{LHCb}
\bibliography{standard,SL,LHCb-PAPER,LHCb-CONF,LHCb-DP,LHCb-TDR,HFLAV-SMpredictions}
\newpage
\centerline
{\large\bf LHCb collaboration}
\begin
{flushleft}
\small
R.~Aaij$^{32}$\lhcborcid{0000-0003-0533-1952},
A.S.W.~Abdelmotteleb$^{50}$\lhcborcid{0000-0001-7905-0542},
C.~Abellan~Beteta$^{44}$,
F.~Abudin{\'e}n$^{50}$\lhcborcid{0000-0002-6737-3528},
T.~Ackernley$^{54}$\lhcborcid{0000-0002-5951-3498},
B.~Adeva$^{40}$\lhcborcid{0000-0001-9756-3712},
M.~Adinolfi$^{48}$\lhcborcid{0000-0002-1326-1264},
P.~Adlarson$^{77}$\lhcborcid{0000-0001-6280-3851},
H.~Afsharnia$^{9}$,
C.~Agapopoulou$^{13}$\lhcborcid{0000-0002-2368-0147},
C.A.~Aidala$^{78}$\lhcborcid{0000-0001-9540-4988}
% ...

{\footnotesize \it

% ...
$^{9}$Universit{\'e} Clermont Auvergne, CNRS/IN2P3, LPC, Clermont-Ferrand, France\\
% ...
$^{13}$LPNHE, Sorbonne Universit{\'e}, Paris Diderot Sorbonne Paris Cit{\'e}, CNRS/IN2P3, Paris, France\\
% ...
$^{32}$Nikhef National Institute for Subatomic Physics, Amsterdam, Netherlands\\
% ...
$^{40}$Instituto Galego de F{\'\i}sica de Altas Enerx{\'\i}as (IGFAE), Universidade de Santiago de Compostela, Santiago de Compostela, Spain\\
% ...
$^{44}$Physik-Institut, Universit{\"a}t Z{\"u}rich, Z{\"u}rich, Switzerland\\
% ...
$^{48}$H.H. Wills Physics Laboratory, University of Bristol, Bristol, United Kingdom\\
% ...
$^{50}$Department of Physics, University of Warwick, Coventry, United Kingdom\\
% ...
$^{53}$School of Physics and Astronomy, University of Glasgow, Glasgow, United Kingdom\\
$^{54}$Oliver Lodge Laboratory, University of Liverpool, Liverpool, United Kingdom\\
% ...
$^{62}$Syracuse University, Syracuse, NY, United States\\
% ...
$^{77}$Department of Physics and Astronomy, Uppsala University, Uppsala, Sweden, associated to $^{53}$\\
$^{78}$University of Michigan, Ann Arbor, MI, United States, associated to $^{62}$\\
\bigskip
% ...
}
\end{flushleft}
\newpage

\section{Detailed signal region fit projections}
\begin{figure*}[htbp]
\includegraphics[width=0.80\textwidth]{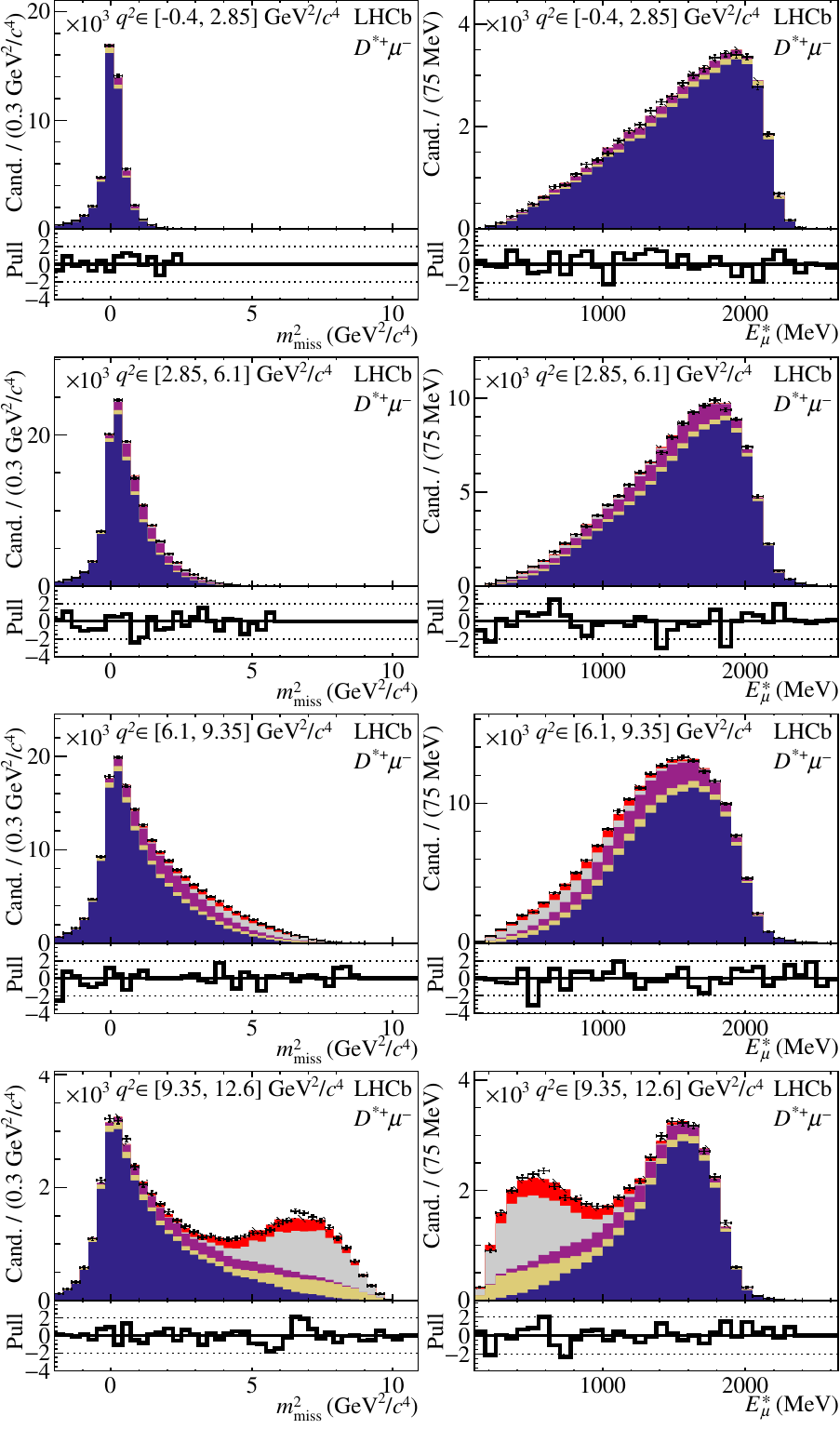}
\raisebox{10.0cm}{\includegraphics[trim=0.5cm 0cm 0cm 0cm, clip=True, width=0.18\textwidth]{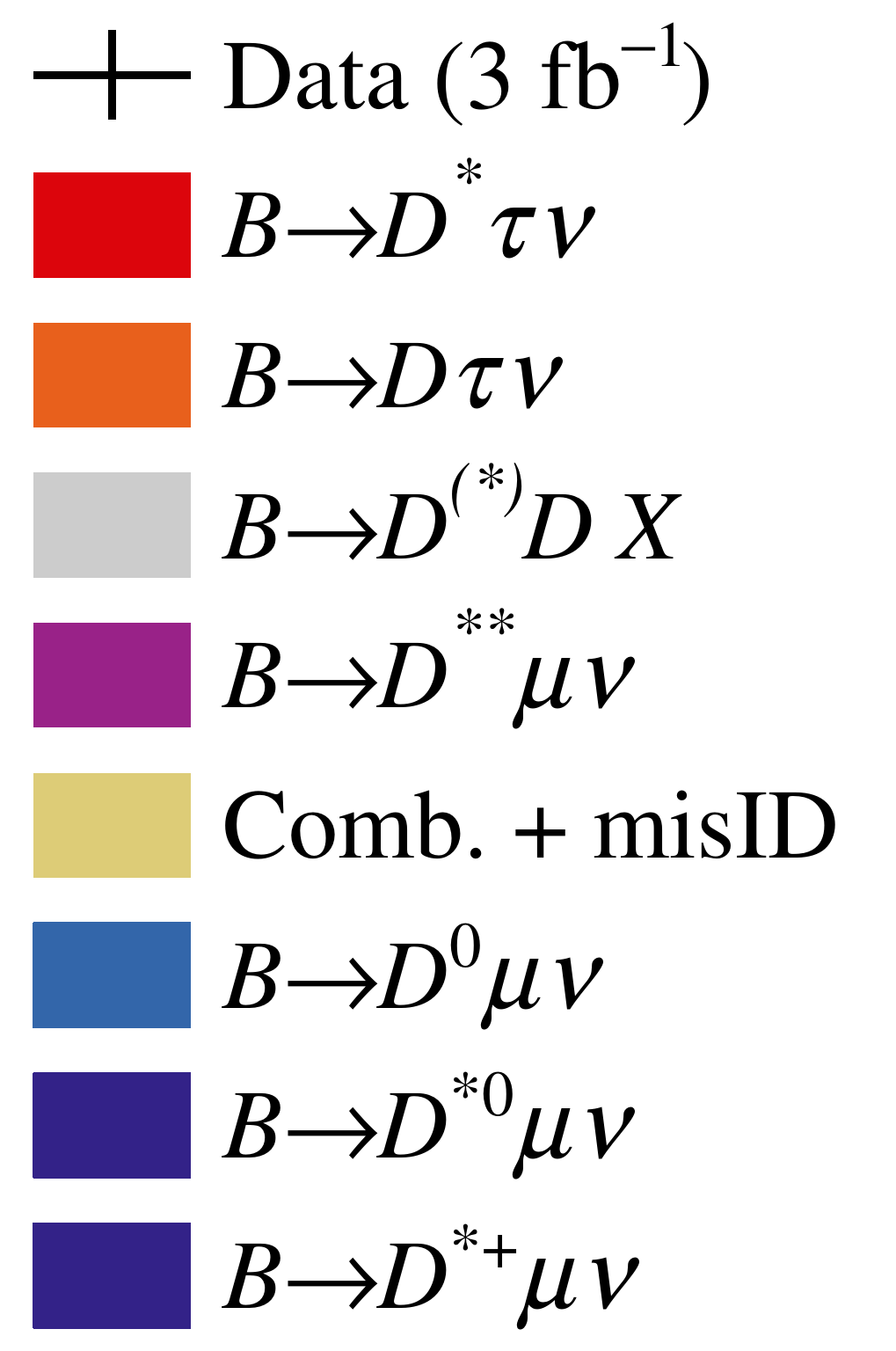}}
\caption{Distributions of (left) $m^2_{\rm miss}$ and (right) \El in each \qq bin of the \Dstarp\mun signal data, overlaid with projections of the fit model. Below each panel differences between the data and fit are shown, normalized by the summed statistical uncertainty of the data and simulation samples.}
\label{fig:dstmuq2bins}
\end{figure*}

\begin{figure*}[htbp]
\centering
\includegraphics[width=0.8\textwidth]{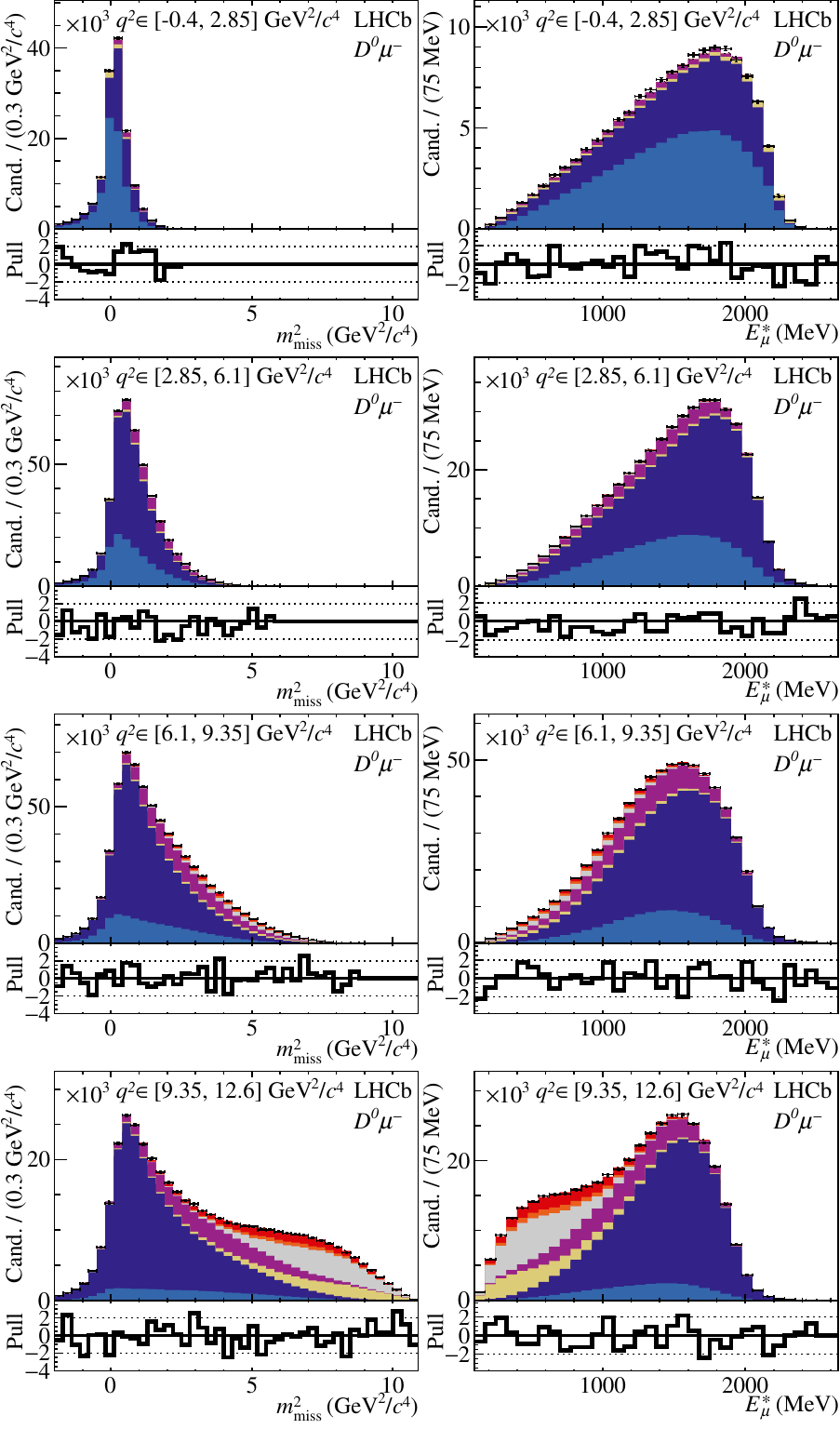}
\raisebox{10.0cm}{\includegraphics[trim=0.5cm 0cm 0cm 0cm, clip=True, width=0.18\textwidth]{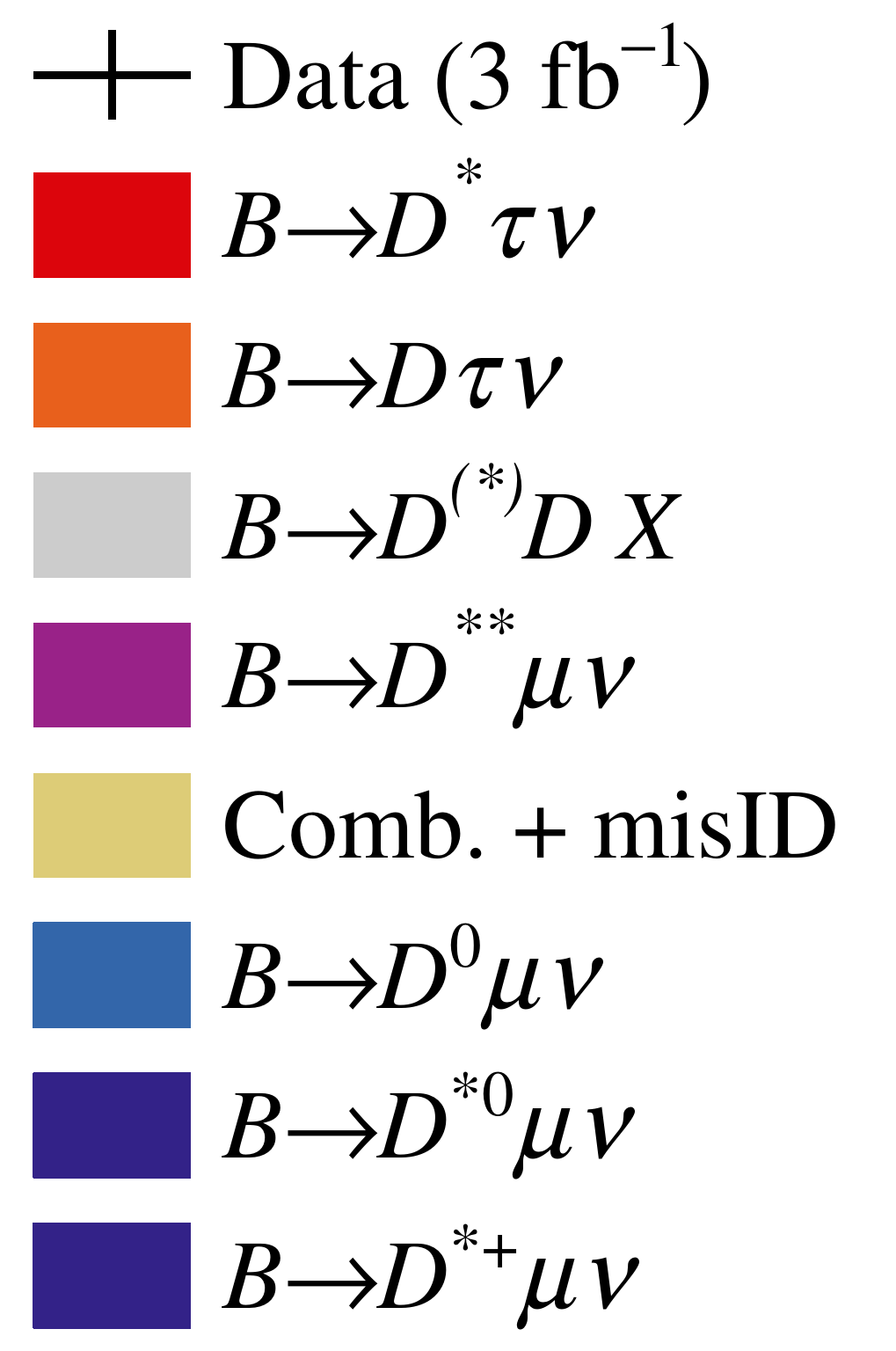}}
\caption{Distributions of (left) $m^2_{\rm miss}$ and (right) \El in each \qq bin of the \Dz\mun signal data, overlaid with projections of the fit model. Below each panel differences between the data and fit are shown, normalized by the summed statistical uncertainty of the data and simulation samples.}
\label{fig:dzmuq2bins}
\end{figure*}

For compactness, Fig.~\ref{fig:theFit} shows only the highest \qq region of the signal fit, where the signal purity is largest. The other \qq bins provide strong constraints on the non-signal contributions to the fit, as well as containing a substantial amount of signal themselves, albeit at a lower purity. 
Figures~\ref{fig:dstmuq2bins} and~\ref{fig:dzmuq2bins} show \mmsq and \El projections of the signal region of the fit for all \qq bins in order to provide a complete picture of the fit quality and completeness of the model. 
\clearpage
\section{Control region fit projections}
\begin{figure*}[htbp]
\centering
\begin{minipage}{0.98\textwidth}
\centering
\includegraphics[width=0.95\textwidth]{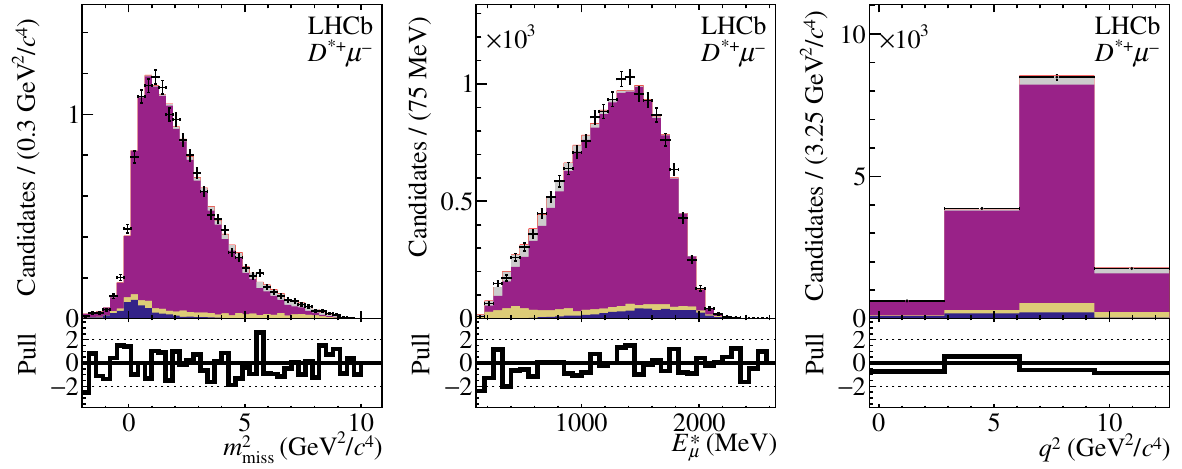}\\
\includegraphics[trim=0cm 0cm 0cm 0cm, clip=True, width=0.8\textwidth]{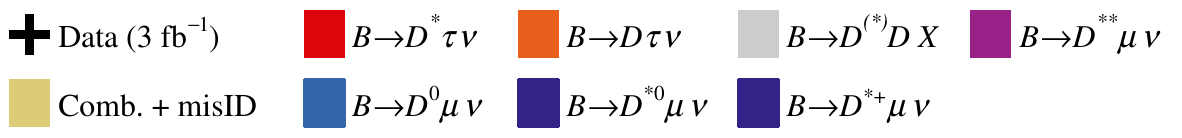}\vspace{8mm}
\end{minipage}
\begin{minipage}{0.98\textwidth}
\centering
\includegraphics[width=0.95\textwidth]{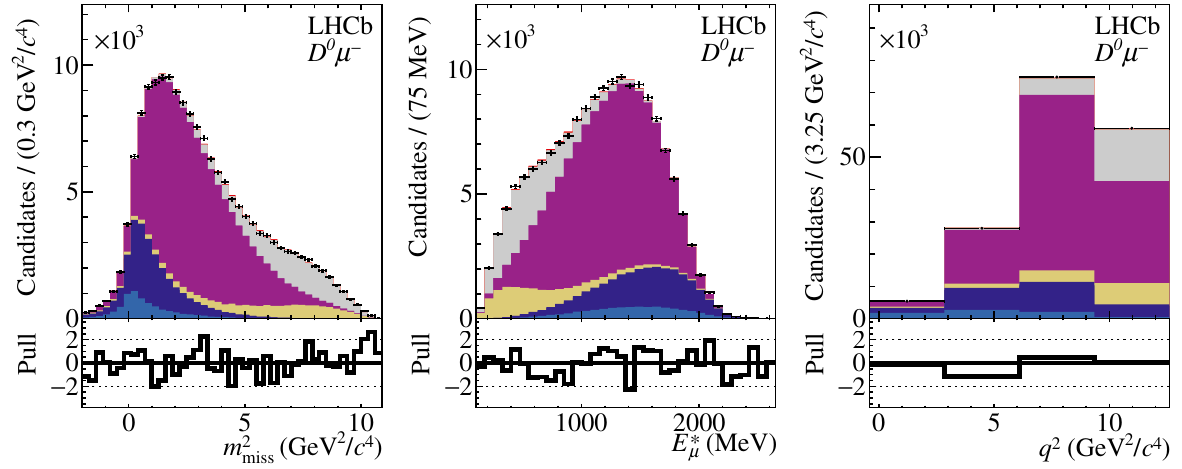}\\
\includegraphics[width=0.8\textwidth]{figs/FigB3b.pdf}
\end{minipage}
\caption{Projections of the fit in (left) missing mass squared, (center) lepton energy, and (right) \qq in the (top) $\Dstar\mun$ and (bottom) $\Dz\mun$ region with exactly one extra pion consistent with the \B vertex.}
\label{fig:supp:1OS}
\end{figure*}

\begin{figure*}[htbp]
\centering
\begin{minipage}{0.98\textwidth}
\centering
\includegraphics[width=0.95\textwidth]{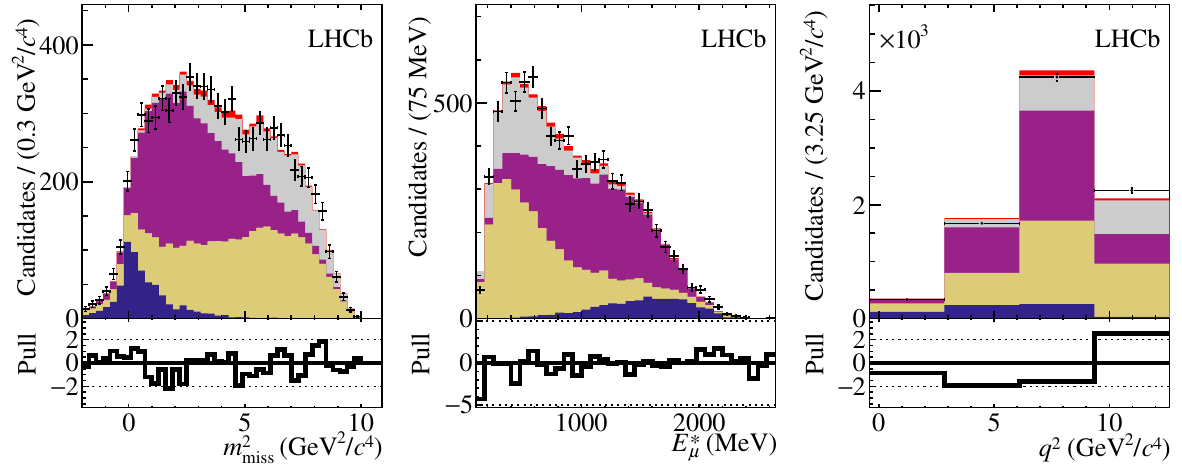}
\includegraphics[width=0.8\textwidth]{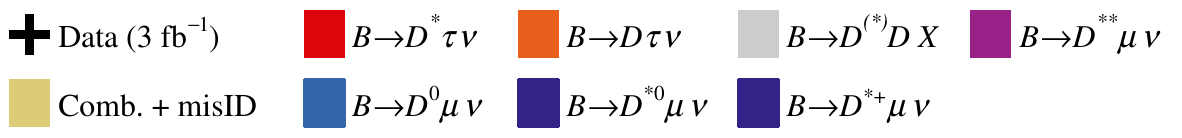}\vspace{8mm}
\end{minipage}
\begin{minipage}{0.98\textwidth}
\centering
\includegraphics[width=0.95\textwidth]{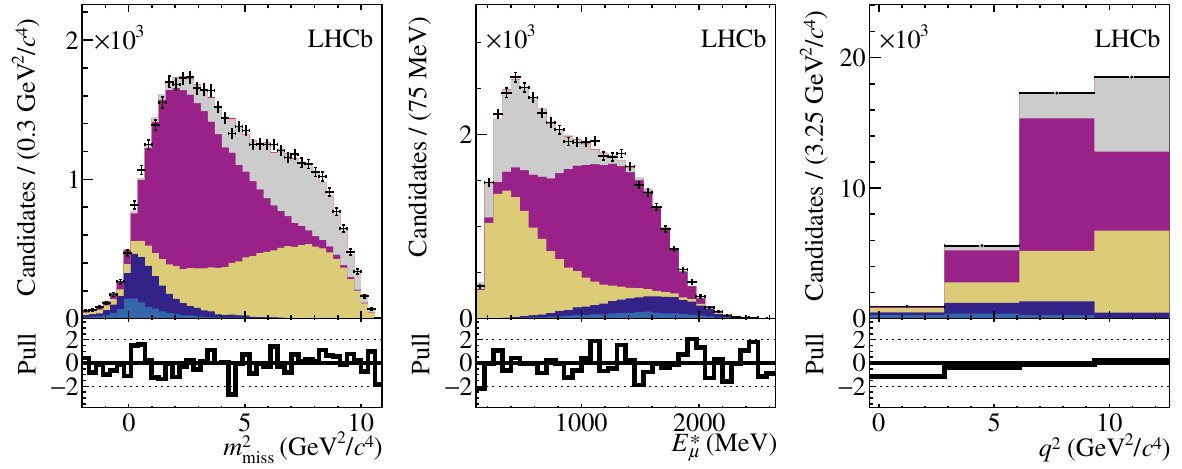}
\includegraphics[width=0.8\textwidth]{figs/FigB4b.pdf}
\end{minipage}
\caption[fit projections in the \Dz ISO region]{Projections of the fit in (left) missing mass squared, (center) lepton energy, and (right) \qq in the (top) $\Dstar\mun$ and (bottom) $\Dz\mun$ region with exactly two extra opposite-sign pions consistent with the \B vertex.}
\label{fig:supp:2OS}
\end{figure*}

\begin{figure*}[htbp]
\centering
\begin{minipage}{0.98\textwidth}
\centering
\includegraphics[width=0.95\textwidth]{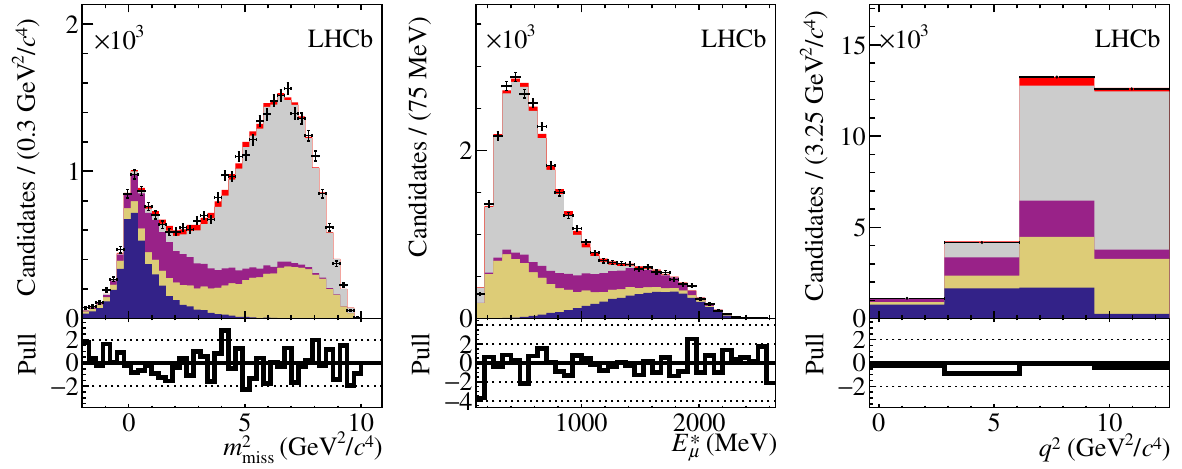}
\includegraphics[width=0.8\textwidth]{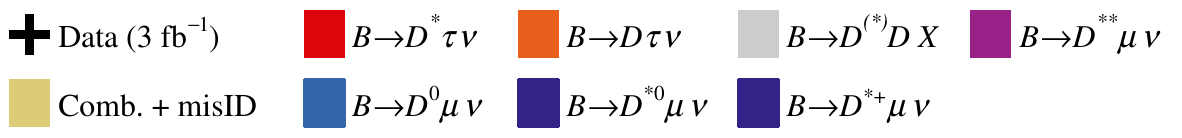}\vspace{8mm}
\end{minipage} 

\begin{minipage}{0.98\textwidth}
\centering
\includegraphics[width=0.95\textwidth]{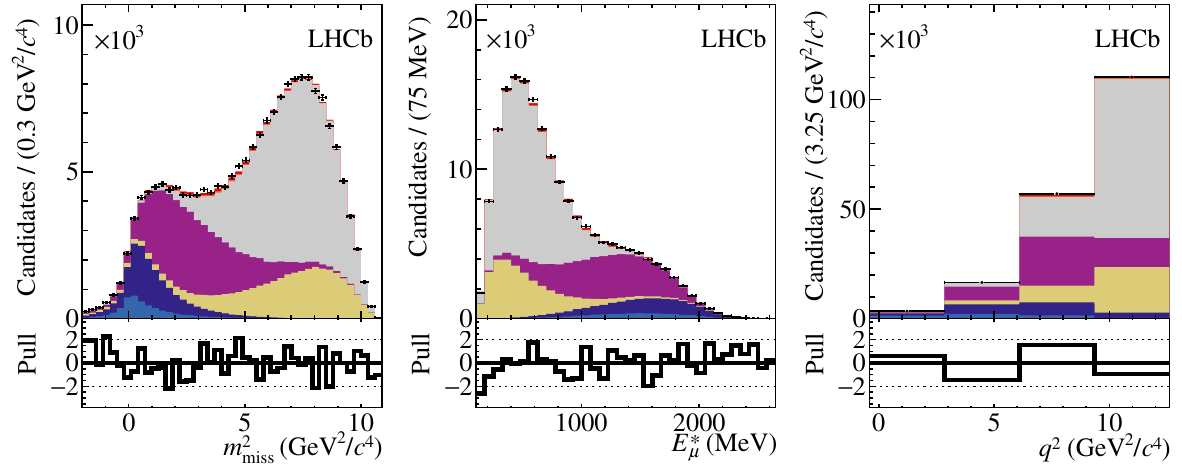}
\includegraphics[width=0.8\textwidth]{figs/FigB5b.pdf}
\end{minipage}
\caption[fit projections in the \Dz ISO region]{Projections of the fit in (left) missing mass squared, (center) lepton energy, and (right) \qq in the (top) $\Dstar\mun$ and (bottom) $\Dz\mun$ region with at least one kaon of either sign and no restrictions on the number of additional tracks consistent with the \B vertex.}
\label{fig:supp:DD}
\end{figure*}

As described in the text, the control regions selected to be fit simultaneously with the signal region are chosen to provide strong constraints on the modeling of the various parameters that determine the shapes of the backgrounds to the signal decays. These regions are fit using the same model as the signal region, but with templates derived from each of the respective selection regions. Figure~\ref{fig:supp:1OS} shows the region with exactly one additional charge-correlated \pion candidate, Fig.~\ref{fig:supp:2OS} shows the region with exactly two additional \pion candidates with opposite charge, and Fig.~\ref{fig:supp:DD} shows the region selected by the presence of a charged \kaon candidate.

\clearpage

\section{Determination of non-muon background using data}
\label{appendix:misID}

To model the contribution of misID background, 
a control sample of $\Dz$ or $\Dstarp$ candidates paired with an extra track is used, where the combinations 
pass all the analysis selection criteria but the extra track taking the place of the muon has no associated segment in the muon system.
Two techniques are used, corresponding with the two fitters developed for this analysis, with the goal of weighting this control sample to produce a model of the misidentified backgrounds. 

In the first technique, used in the primary analysis reported here, the extra track in the $D^{*}$-track combination is categorized according to whether it passes particle identification (PID) criteria for $\pi$, $K$, $p$, or $e$, or whether it fails all four. The distribution of each identified particle species in a given kinematic bin may be fit using templates derived from large PID control samples. From these fits, {\it sWeights}~\cite{Pivk:2004ty} are produced to statistically separate the properties of the contribution from $\pi$, $K$, $p$, $e$ or misreconstructed tracks (``ghosts"). The {\it sWeights} are combined with muon misidentification rates measured in PID control samples to obtain per-track weights that, summed over all categories and all tracks, give a model of the misID background given the information in the $D^{*}$-track and PID control samples.

In the second technique, first employed in Ref.~\cite{LHCb-PAPER-2017-035}, the number of candidates passing each criteria as a function of
track kinematics is used in conjunction with large PID calibration samples to unfold \cite{RooUnfold} the estimated true fraction of each species $h$ ($h=$\pion, \kaon, \proton, or \electron, or misreconstructed ``ghost" tracks) in a given kinematic bin. These unfolded numbers $f_h$ are
taken as prior probabilities that any given track's true identity is one of the five species. The efficiencies for the category selections, measured in control data, are taken as conditional probabilities that a track of one true identity $h$ is categorized as identity $h^{\prime}$, $P(h'|h)$. These two pieces of information may be combined using Bayes's rule to obtain the probability $P(h|h')$ that a track identified as $h'$ has a given true identity $h$. These posterior probabilities are combined with muon misidentification rates from
the PID calibration samples to produce a final weight for each track summed over the five hypotheses.

Due to the requirement that the tracks used in the misID modeling techniques fail muon identification criteria, there is no component of $\pi$ or $K$ decays in flight (\pip or $\Kp \to \mup \neum$) in the $D^{(*)}$-track sample used above. However, these are known to represent a significant contribution of the hadron to muon misidentification rate \cite{LHCb-DP-2013-001}. 
These decays in flight result in a mismeasurement of the ``muon'' momentum due to the deflection of the muon decay product with respect to the original $\pi$ or $K$ trajectory. This mismeasurement of the $\pi$ or $K$ momentum is included in the misID background model using a smearing derived from simulation.
A simulated sample of $\Dz \to \Km \pip $ is used, with muon identification criteria applied to either the kaon or the pion identical to the selection used in data.
The mismeasurements of the track kinematics in this simulation sample are used  to build an ensemble of possible momentum modifications, which are randomly sampled to build an anzatz for smearing the momenta of the $D^{(*)}$-track candidates. The exact amount of smearing is allowed to vary in the fit to the data to account for differences between this simulation and the misID background.

Both techniques for modeling the misID background are validated by fitting data samples in which the
muon candidate passes initial muon identification criteria, but fails to pass the custom multivariate muon identification developed for
this analysis. 
This sample contains a higher number of misidentified muon candidates than are present in the signal samples, and tends to have a larger average momentum mismeasurement.
The two techniques show excellent agreement with this sample once the smearing is applied.

\section{Agreement with previous measurement}
\label{appendix:past}
The $\Dstarp \mun$ sample in this measurement overlaps completely with that previously analysed in Ref.~\cite{LHCb-PAPER-2015-025}. To assess the level of agreement with the value of \RDst in Ref.~\cite{LHCb-PAPER-2015-025}, a fit is performed to the $\Dstarp \mun$ signal sample alone, with all shape parameters fixed to their best fit values from the default fit.
The uncorrelated systematic uncertainty between the value of \RDst from this fit and that in Ref.~\cite{LHCb-PAPER-2015-025} is estimated from the quadratic difference in individual systematic uncertainties (excluding the misID unfolding, where $100\%$ of the previous uncertainty is treated as uncorrelated, and the simulation statistical uncertainty where $1/\sqrt{2}$ times the present uncertainty is assigned). The quadrature difference in statistical uncertainties is also included as the uncorrelated effect of the reduced statistical power of the data due to the more stringent muon identification requirements used here.  This procedure gives an estimated uncorrelated uncertainty of 0.026 in \RDst.
The measured value of \RDst in this fit is 0.293, in $1.6\sigma$ agreement with Ref.~\cite{LHCb-PAPER-2015-025}.


\end{document}